\documentclass[12pt]{article}
\usepackage{amsmath,amssymb,exscale,cancel}
\usepackage{slashed}
\usepackage{graphicx}
\usepackage{epsfig}
\usepackage{multicol}
\usepackage{color}
\usepackage{mathrsfs}
\usepackage{blindtext}
 \usepackage{fancyhdr}
\usepackage{hyperref}
\usepackage{cite}
\usepackage{mathtools}

\usepackage{floatrow}

\usepackage{graphicx}
\usepackage{sidecap}

\usepackage[latin1]{inputenc} 

\usepackage{multicol}  
 
 \usepackage{titlesec}
 
 \usepackage{rotating,slashed,xcolor,amsfonts,expdlist,charter}

\numberwithin{equation}{section}

\usepackage{xcolor}
\usepackage{sectsty}

\usepackage{mdframed}
\usepackage{titletoc}

\numberwithin{equation}{section}

\usepackage{xcolor}
\usepackage{sectsty}

\usepackage{mdframed}
\usepackage{titletoc}

\definecolor{secnum}{RGB}{13,151,225}
\definecolor{ptcbackground}{RGB}{212,237,252}
\definecolor{ptctitle}{RGB}{0,177,235}

\titlecontents{lsection}
  [5.8em]{\sffamily}
  {\color{secnum}\contentslabel{2.3em}\normalcolor}{}
  {\titlerule*[1000pc]{.}\contentspage\\\hspace*{-5.8em}\vspace*{5pt}%
    \color{white}\rule{\dimexpr\textwidth-15.5pt\relax}{1pt}}

\usepackage{hyperref}
\hypersetup{colorlinks,bookmarksopen,bookmarksnumbered,citecolor=rossos,
linkcolor=redy,pdfstartview=FitH,urlcolor=green-go}
\usepackage{slashed}

\definecolor{blus}{cmyk}{1,0.9,0,0.1}
\definecolor{verdes}{cmyk}{0.99,0,0.59,0.65}
\definecolor{rossos}{cmyk}{0,1,1,0.55}
\definecolor{redy}{cmyk}{0,1,1,0.7}
\definecolor{greeny}{cmyk}{0.99,0,0.59,0.98}
\definecolor{green-go}{cmyk}{0.79,0,0.59,0.5}

\usepackage{titlesec}

\def\Lag{\mathscr{L}}

\def\tt{{\tiny \times}}

\newcommand{\beq}{\begin{equation}}
\newcommand{\eeq}{\end{equation}}

\def\hhref#1{\href{http://arxiv.org/abs/#1}{arXiv:#1}} 

 \def\Lag{\mathscr{L}}
 
\newcommand{\tmtextbf}[1]{{\bfseries{#1}}}
\newcommand{\tmtextrm}[1]{{\rmfamily{#1}}}
\newcommand{\bp}{\bar M_P}

\def\be{\begin{equation}}
\def\ee{\end{equation}}
\def\ba{\begin{array} }

\def\bac{\begin{array} {c}}
\def\bacc{\begin{array} {cc}}
\def\baccc{\begin{array} {ccc}}
\def\bacccc{\begin{array} {cccc}}
\def\ea{\end{array}}
\def\bea{\begin{eqnarray}}
\def\eea{\end{eqnarray}}

\definecolor{red}{rgb}{1,0,0}

\def\psl{\hbox{\hbox{${p}$}}\kern-1.9mm{\hbox{${/}$}}}
\def\dsl{\hbox{\hbox{${\partial}$}}\kern-2.2mm{\hbox{${/}$}}}
\def\Dsl{\hbox{\hbox{${D}$}}\kern-2.6mm{\hbox{${/}$}}}

\def\Lag{\mathscr{L}}

\newcommand{\gappeq}{{\rlap{{\raise}.5ex\text{\ensuremath{>}}}{{\lower}.5ex\text{\ensuremath{\sim}}}}}
\newcommand{\lappeq}{{\rlap{{\raise}.5ex\text{\ensuremath{<}}}{{\lower}.5ex\text{\ensuremath{\sim}}}}}
\newcommand{\I}{\tmtextrm{1{\kern}-.24em l}}

\begin{document}
\topmargin -1.0cm
\oddsidemargin 0.9cm
\evensidemargin -0.5cm

{\vspace{-1cm}}
\begin{center}

\vspace{-1cm}

 {\Huge \tmtextbf{ 
\color{rossos} \hspace{-0.9cm}  Model-Independent \\
Radiative Symmetry Breaking \\ and Gravitational Waves \hspace{-1.6cm}}} {\vspace{.5cm}}\\


\vspace{1.3cm}

{\large  {\bf Alberto Salvio }
{\em  
\vspace{.4cm}

 Physics Department, University of Rome Tor Vergata, \\ 
via della Ricerca Scientifica, I-00133 Rome, Italy\\

\vspace{0.6cm}

I. N. F. N. -  Rome Tor Vergata,\\
via della Ricerca Scientifica, I-00133 Rome, Italy\\

\vspace{.4cm}


\vspace{0.4cm}

\vspace{0.2cm}

 \vspace{0.5cm}
}

\vspace{.3cm}

}
\vspace{0.cm}

\end{center}

%
%
\noindent ---------------------------------------------------------------------------------------------------------------------------------
\begin{center}
{\bf \large Abstract}
\end{center}
\noindent Models where symmetries are predominantly broken (and masses are then generated) through radiative corrections typically produce strong first-order phase transitions with a period of supercooling, when the temperature dropped by several orders of magnitude. Here it is shown that a model-independent description of these phenomena and the consequent production of potentially observable  gravitational waves is possible in terms of few parameters (which are computable once the model is specified) if enough supercooling occurred. It is explicitly found how large the supercooling should be in terms of those parameters, in order for the model-independent description to be valid. It is also explained how to systematically improve the accuracy of such description by computing higher-order corrections in an expansion in powers of a small quantity, which is a function of the above-mentioned parameters. Furthermore, the corresponding gravitational wave spectrum is compared with the existing experimental results from the latest observing run of LIGO and VIRGO and the expected sensitivities of future gravitational wave experiments to find regions of the parameter space that are either ruled out or can lead to a future detection.

\vspace{0.7cm}

\noindent---------------------------------------------------------------------------------------------------------------------------------

  \vspace{-0.9cm}
  
  \newpage 
\tableofcontents

\noindent

\vspace{0.5cm}

\section{Introduction}\label{Introduction}

Current and future gravitational wave (GW) detectors provide us with precious information not only regarding astrophysical systems, such as black holes and other compact objects~\cite{Abbott:2016blz,TheLIGOScientific:2016wyq1,LIGOScientific:2017ync}, but also in relation to particle physics and the corresponding phenomenology. In particular, they can probe high energies, even higher than those accessible at particle accelerators. A classic example is the fact that strong first-order phase transitions, which are predicted by some Standard Model (SM) high-energy extensions, can generate a background of GWs that may be observable (see Ref.~\cite{Maggiore:2018sht} for a textbook introduction). Remarkably, the possible observation of GWs due to a first-order phase transition (PT) would be a clear signal of new physics because the SM does not feature this type of PTs.

The phases that are separated by a PT can  have different symmetry properties and some global and/or gauge symmetries can be recovered at high temperatures~\cite{Weinberg:1974hy}, higher than the critical temperature. In the low-temperature limit these symmetries can be broken by the Higgs mechanism, like in the SM. However, there are other options beyond the SM. One of the most famous alternative is the possibility of breaking some symmetries  (and correspondingly generate masses) through radiative (i.e. perturbative loop) corrections. The seminal work on such radiative symmetry breaking (RSB) is Ref.~\cite{Coleman:1973jx} by Coleman and E.~Weinberg, which considered a simple toy model (see also Ref.~\cite{Levi:2022bzt}
for a recent analysis). The Coleman-Weinberg paper was later extended to a more general field theory by Gildener and S.~Weinberg~\cite{Gildener:1976ih}. An important feature of the RSB scenario is approximate scale invariance; radiative corrections generically break this symmetry, but the breaking is small as long as the theory is perturbative.  

Many examples of  RSB models featuring a strong first-order PT and predicting potentially observable GWs  are known. These range from electroweak (EW) symmetry breaking~\cite{Espinosa:2008kw,Farzinnia:2014yqa,Sannino:2015wka,Marzola:2017jzl,Brdar:2019qut,Kierkla:2022odc} to grand unified models~\cite{Huang:2020bbe}, passing through, for example, Peccei-Quinn~\cite{Peccei:1977hh}   symmetry breaking~\cite{DelleRose:2019pgi,VonHarling:2019rgb,Salvio:2020prd,Ghoshal:2020vud} and the seesaw mechanism~\cite{Brdar:2018num,Kubo:2020fdd}, see Ref.~\cite{Salvio:2020axm} for a review. 

The vast number of models where RSB leads to a strong first-order PT and to potentially observable GWs suggests that there may be a model-independent description of these phenomena. Since in works of this type there is typically some degree of supercooling (the temperature drops by several orders of magnitude below the critical temperature before the PT really takes place in the expanding universe), it is natural to conjecture that this phenomenon may play an important and general role. The main objective of  this work is to find out if a model-independent proof of the presence of a strong first-order PT exists and, if so, whether there is also a model-independent description of such PT and the consequent GW spectrum.  In order to proceed perturbatively, this study also implies an analysis of the validity of the loop and derivative expansions in the general RSB scenario.  Another purpose of the present work is to establish whether supercooling is a key and general ingredient for a first-order PT and observable GWs in the RSB scenario.

The consequent theoretical GW spectrum can then be compared, without any need to specify a model, with the constraints and the expected sensitivities of current and future GW experiments. These include  ground-based interferometers, such as the advanced  Laser Interferometer Gravitational-Wave Observatory (LIGO)~\cite{Harry:2010zz,TheLIGOScientific:2014jea}, Advanced Virgo~\cite{VIRGO:2014yos}, Cosmic Explorer (CE)~\cite{Evans:2016mbw,Reitze:2019iox} and 
Einstein Telescope (ET)~\cite{Punturo:2010zz, Hild:2010id, Sathyaprakash:2012jk})
as well as  space-based interferometers, like the Big Bang Observer (BBO)~\cite{Crowder:2005nr, Corbin:2005ny, Harry:2006fi}, the Deci-hertz Interferometer Gravitational wave Observatory (DECIGO)~\cite{Seto:2001qf, Kawamura:2006},
 the Laser Interferometer Space Antenna (LISA)~\cite{Audley:2017drz}, etc.

 An important advantage of the above-mentioned model-independent analysis is of course the fact that one could quickly have information on any given setup once a model is specified, without repeating every time the full analysis of the PT and the GW spectrum. 
 
The paper is structured as follows.
\begin{itemize}
\item In Sec.~\ref{theoretical framework} the general theoretical framework of RSB,  where the  masses are mostly generated radiatively, is analysed. A general number of scalar, fermion and vector fields with arbitrary couplings is considered so that the analysis can be model-independent. In this framework we study the one-loop quantum effective potential and the necessary generation of the EW scale. To reach the main objective of the paper, the corresponding one-loop thermal potential is also studied there.
\item In Sec.~\ref{PT} we study the PT associated with RSB and investigate its nature in full generality. There, the role of supercooling is also considered, again without specifying the model.
\item Sec.~\ref{Gravitational Waves} is then dedicated to the study of the possible GWs produced by the PT in the general RSB scenario and the comparison with the current and future GW detectors.
\item To obtain a model-independent description of any phenomenon it is necessary to perform some approximations. The accuracy of these approximations and how to improve them is then discussed in Sec.~\ref{impro}.
\item In Sec.~\ref{Conclusions} a detailed summary of the results of this work as well as the final conclusions are provided. 
\end{itemize}

\section{General theoretical framework}\label{theoretical framework}
Since in the RSB scenario the masses are mostly generated through radiative corrections, we start from the most general no-scale matter Lagrangian describing  the interactions between the matter fields:
\be \label{eq:Lmatterns}
\Lag^{\rm ns}_{\rm matter} =  
- \frac14 F_{\mu\nu}^AF^{A\mu\nu} + \frac{D_\mu \phi_a \, D^\mu \phi_a}{2}  + \bar\psi_j i\slashed{D} \psi_j  - \frac12 (Y^a_{ij} \psi_i\psi_j \phi_a + \hbox{h.c.}) 
- V_{\rm ns}(\phi), 
\ee 
while gravity is assumed to be described by standard Einstein's theory at the energies that are relevant for this work.
Here we consider generic numbers 
of real scalars $\phi_a$,   Weyl fermions $\psi_j$ and vectors $V^A_\mu$ (with field strength $F_{\mu\nu}^A$), respectively. The $V^A_\mu$ are gauge fields
and allow us to construct the covariant derivatives
$$D_\mu \phi_a = \partial_\mu \phi_a+ i \theta^A_{ab} V^A_\mu \phi_b, \qquad D_\mu\psi_j = \partial_\mu \psi_j + i t^A_{jk}V^A_\mu\psi_k, $$ 
where $\theta^A$ and $t^A$ are the generators of the (internal) gauge group in the scalar and fermion representations, respectively. Note that, since we are working with real scalars, the Hermitian matrices $\theta_A$ are purely imaginary and antisymmetric. The gauge couplings are contained in the $\theta^A$ and $t^A$.
Also, the $Y^a_{ij}$  are the Yukawa couplings  and $V_{\rm ns}(\phi)$ is the no-scale potential,
\be V_{\rm ns}(\phi)= \frac{\lambda_{abcd}}{4!} \phi_a\phi_b\phi_c\phi_d, \label{Vns}\ee
($\lambda_{abcd}$ are the quartic couplings). We take $\lambda_{abcd}$ totally symmetric with respect to the exchange of its indices $abcd$ without loss of generality. In~(\ref{eq:Lmatterns}) all terms are contracted in a gauge-invariant way.

 \subsection{Radiative symmetry breaking}\label{CWGW}

  In the RSB mechanism the mass scales emerge radiatively from loops in a way we discuss now.
   The basic idea is that, since at quantum level the couplings depend on the RG energy $\mu$, there may be some specific energy 
at which the potential  in Eq.~(\ref{Vns}) develops a flat direction. Such flat direction can be written as $\phi_a = \nu_a \chi$, where $ \nu_a$ are the components of a unit vector $\nu$, i.e.~$ \nu_a  \nu_a =1$, and $\chi$ is a single scalar field, 
which parameterizes this direction. 
Therefore, after renormalization, the RG-improved potential $V$ along the flat direction reads
\be V(\chi) = \frac{\lambda_\chi (\mu)}{4}\chi^4,  \label{Vvarphi}\ee 
where 
\be \lambda_\chi(\mu) \equiv\frac1{3!} \lambda_{abcd}(\mu) \nu_a \nu_b \nu_c \nu_d. \label{lambdaphi}\ee
Having a flat direction along $\nu$ for $\mu$ equal to some specific value $\tilde\mu$ means 
 \be \lambda_\chi(\tilde\mu)\equiv\lambda_{abcd}(\tilde\mu) \nu_a \nu_b \nu_c \nu_d=0. \ee
 
  Besides the potential in~(\ref{Vvarphi}), quantum loop corrections also generate other terms $V_1+V_2+...$, where $V_i$ represents the $i$-loop contribution. The explicit expression of $V_1$ is well known. Here we can recover it, without specifying the details of the underlying theory, by recalling that the effective potential does not depend on $\mu$. Indeed, the renormalization changes the couplings, the masses and  the fields, but leaves the Lagrangian (and in particular the potential) invariant. So we can write
 \be \mu \frac{dV_q}{d\mu} =0, \qquad \mbox{where}\quad V_q\equiv V+V_1+V_2+...\, .\ee
 Using~(\ref{Vvarphi}), the solution of this equation at the one-loop level is
 \be V_q=\frac{\lambda_\chi (\mu)}{4}\chi^4+ \frac{\beta_{\lambda_\chi}}4\left(\log\frac{\chi}{\mu}+a_s\right)\chi^4,  \ee
 where 
 \be \beta_{\lambda_\chi} \equiv  \mu\frac{d\lambda_{\chi}}{d\mu} \ee
 is the beta function of $\lambda_\chi$ 
 and  $a_s$ is a renormalization-scheme-dependent quantity. Setting now $\mu=\tilde\mu$ where $\lambda_\chi=0$, one obtains
 \be V_q(\chi) = \frac{\bar \beta}4\left(\log\frac{\chi}{\chi_0}-\frac14\right)\chi^4,\label{CWpot}\ee
where
 \be \bar\beta \equiv \left[\beta_{\lambda_\chi} \right]_{\mu=\tilde\mu}, \qquad \chi_0\equiv \frac{\tilde\mu}{e^{1/4+a_s}}.\ee
 Note that the renormalization-scheme-dependent $a_s$ has been absorbed in the scale $\chi_0$.
 We see that the flat direction acquires some steepness at loop level.  The field value $\chi_0$ is a stationary point of $V_q$.
Moreover, $\chi_0$ is a point of minimum when $\bar\beta>0$. Therefore, when the conditions 
  \beq\left\{
\begin{array}{rcll}
\lambda_\chi(\tilde\mu)  &=& 0 & \hbox{(flat direction),}\\
 & & \\ 
\beta_{\lambda_\chi}(\tilde\mu)  &>& 0 & \hbox{(minimum condition),}
\end{array}\right.
\label{eq:CWgen}
\eeq
 are satisfied quantum corrections generate a minimum of the potential at a non-vanishing value of $\chi$, that is  $\chi_0$. In that case $\chi_0$ is the (radiatively induced) zero-temperature vacuum expectation value of $\chi$ and the fluctuations of $\chi$ around $\chi_0$ have squared mass $m^2_\chi=\bar\beta\chi_0^2$. 
 
  This non-trivial minimum can generically break global and/or local symmetries and thus generate the  particle masses, with $\chi_0$ playing the role of the symmetry breaking scale.  Consider for example a term in the Lagrangian density $\Lag$ of the form 
 \be \Lag_{\chi h}\equiv \frac12 \lambda_{ab} \phi_a\phi_b |{\mathcal H}|^2,\label{LvarphiH}\ee 
 where ${\mathcal H}$ is the Standard Model (SM) Higgs doublet and the $\lambda_{ab}$ are some of the quartic couplings. RG-improving and setting $\mu=\tilde\mu$ and $\phi$ along the flat direction, $\nu$,
 \be  \Lag_{\chi {\mathcal H}} = \frac12 \lambda_{\chi h}(\tilde\mu) \chi^2 |{\mathcal H}|^2,\label{portal} \ee
 where 
 \be \lambda_{\chi h}(\mu) \equiv  \lambda_{ab}(\mu) \nu_a\nu_b.\ee
 Thus, by evaluating this term at the minimum $\chi=\chi_0$ we obtain the Higgs squared mass parameter
  \be \mu_h^2 = \frac12\lambda _{\chi h}(\tilde\mu) \chi_0^2. \ee
 In order to provide a mass to the SM elementary particles, we need $\mu_h^2>0$, namely we have the additional condition
 \be \lambda _{\chi h}(\tilde\mu) >0\quad  \hbox{(generation of the EW scale)}. \ee

 \subsection{Thermal effective potential}\label{TVeff}
 

 In order to write a general formula for the thermal contribution to the effective potential, $V_{\rm eff}$, we need to write general expressions for the background-dependent masses. 
 
In the scalar sector the elements of the squared-mass matrix are given by the Hessian matrix  of the no-scale classical potential in~(\ref{Vns}):
\be M_{Sab}^2  \equiv \frac{\partial^2V_{\rm ns}}{\partial\phi_a\partial\phi_b} =\frac12 \lambda_{abcd} \phi_c\phi_d, \label{MS20}\ee 
By evaluating this Hessian matrix at the flat direction $\phi =\nu\chi$ we obtain that the scalar squared-mass matrix $M_S^2$ is proportional to $\chi^2$ via some quartic couplings, namely
\be M_{Sab}^2(\chi)=\frac12\lambda_{abcd}\nu_c\nu_d\chi^2. \label{MS2}\ee
Since $M_S^2$ is real and symmetric it can be diagonalized with a real orthogonal matrix, to obtain $M_S^2(\chi)\to$\,\,diag$(...,m_s^2(\chi), ...)$, where the $m_s(\chi)$ are the background-dependent scalar masses, the eigenvalues of $M_S^2(\chi)$. All the  $m_s^2$ must be non-negative to have a classical potential bounded from below. To prove this first note that 
the requirement that the classical potential in~(\ref{Vns}) is bounded from below also implies that potential has the absolute minimum at $\phi=0$  and this minimum vanishes (because no scales are present in~(\ref{Vns})). Also the classical potential is constantly equal to its value at $\phi=0$ along the flat direction otherwise that direction would not be flat. So for a classical potential that is bounded from below and has a flat direction $\nu\chi$
\be V_{\rm ns}(\phi) = V_{\rm ns}(\nu \chi)+\frac{\partial V_{\rm ns}}{\partial\phi_a}(\nu \chi) \delta\phi_a+\frac12\frac{\partial^2V_{\rm ns}}{\partial\phi_a\partial\phi_b}(\nu \chi) \delta\phi_a\delta\phi_b=\frac12\frac{\partial^2V_{\rm ns}}{\partial\phi_a\partial\phi_b}(\nu \chi) \delta\phi_a\delta\phi_b,  \label{ArgPosMS}\ee
where $\delta\phi\equiv \phi-\nu\chi$ is taken here infinitesimal. So if there were negative eigenvalues of $M_{Sab}^2(\chi)$ the classical potential would become smaller than its value at the flat direction, but we have seen that this is not possible for  a bounded-from-below potential. So all $m_s(\chi)$ must be real and, of course, can be taken non-negative.

 This nice property is not shared by theories where symmetry breaking is entirely due to the standard Higgs mechanism, which always requires non-convex regions of the tree-level potential and thus some negative scalar squared masses for some field values.

Let us now turn to the fermion sector. For a given constant field background $\phi$, we choose a fermion basis where $\mu_F \equiv Y^a\phi_a$ (as well as $\mu_F^\dagger$) is diagonal, which can be obtained through an $SU(N_F)$ transformation acting on the fermion fields\footnote{This is known as the complex Autonne-Takagi factorization, see also Ref.~\cite{Youla}.}. The squared-mass matrix is
\be M_F^2 \equiv \mu_F\mu_F^\dagger. \ee 
By evaluating $\phi$ at the flat direction one obtains $\mu_F(\chi) = Y_\nu \chi$, where $Y_\nu\equiv Y^a\nu_a$, and 
\be M_F^2(\chi) =  Y_\nu Y_\nu^\dagger \chi^2. \ee 
In our fermion basis $M_F^2(\chi)=$\,\,diag$(...,m_f^2(\chi), ...)$, where the $m_f(\chi)$ are the background-dependent fermion masses.
Given that $M_F^2$ is the product of $\mu_F$ times its Hermitian conjugate all the $m_f$ are real and, of course, can be taken non-negative.

Finally, in the vector sector the elements of the squared-mass matrix $M^2_V$ are
\be M^2_{VAB}\equiv \phi^T\theta^A\theta^B\phi,\ee
and, evaluating at the flat direction, $M^2_{VAB}(\chi)\equiv \nu^T\theta^A\theta^B\nu \chi^2$.
Since the $\theta^A$ are Hermitian, purely imaginary and antisymmetric, $M^2_V$ is always real, symmetric and non-negatively defined: one can diagonalize $M^2_V$ with a real orthogonal matrix, to obtain $M^2_V\to$
\,\,diag$(...,m_v^2(\chi), ...)$, where the $m_v(\chi)$ are the background-dependent vector masses, the eigenvalues of $M_V^2(\chi)$, and are, just like the $m_s$ and $m_f$, all real. Of course, they can also be taken non-negative.

Including the thermal corrections, the general expression of the effective potential $V_{\rm eff}$ is then 
 (in the Landau gauge and at one-loop level)
\be V_{\rm eff}(\chi,T) = V_q(\chi) +\frac{T^4}{2\pi^2}\left(\sum_b n_b J_B(m_b^2(\chi)/T^2)-2\sum_f J_F(m_f^2(\chi)/T^2)\right)+\Lambda_0,  \label{VeffSumm}  \ee
where $V_q(\chi)$ is given in~(\ref{CWpot}), the sum over $b$ runs over all bosonic degrees of freedom and $n_b=1$ for a scalar degree of freedom (we work with real scalars) and $n_b=3$ for a vector degree of freedom.  In~(\ref{VeffSumm}) the sum over $f$, which runs over the fermion degrees of freedom, is multiplied by 2 because we work with Weyl spinors. Also,  
the thermal functions $J_B$ and $J_F$ are defined by 
\bea \, \hspace{-1cm}J_B(x)\equiv \int_0^\infty dp\, p^2 \log\left(1-e^{-\sqrt{p^2+x}}\right)&=&-\frac{\pi^4}{45}+\frac{\pi^2}{12} x -\frac{\pi}{6} x^{3/2} -\frac{x^2}{32} \ln\left(\frac{x}{a_B}\right) + O(x^3), \label{JBdef}\\
\hspace{-1cm}J_F(x)\equiv \int_0^\infty dp\, p^2 \log\left(1+e^{-\sqrt{p^2+x}}\right)&=& \frac{7\pi^4}{360}-\frac{\pi^2}{24} x -\frac{x^2}{32} \ln\left(\frac{x}{a_F}\right) + O(x^3) \label{JFdef}.\eea
In the equations above we also wrote the expansions of $J_B(x)$ and $J_F(x)$ around $x=0$ modulo terms of order
$x^3$, where $a_B = 16\pi^2 \exp(3/2-2\gamma_E)$, $a_F = \pi^2 \exp(3/2-2\gamma_E)$ and  $\gamma_E$ is the Euler-Mascheroni constant (see Ref.~\cite{Dolan:1973qd} for the derivation of those expansions). In Eq.~(\ref{VeffSumm}) we have included  a constant term $\Lambda_0$ to account for the observed value of the cosmological constant when $\chi$ is set to the value corresponding to the minimum of $V_{\rm eff}$.
This addition does not spoil the argument presented above that shows that for a scale-invariant and bounded-from-below potential with a flat direction the eigenvalues of $M_S^2$ are always non-negative: adding $\Lambda_0$ to $V_{\rm ns}$ just produces an additive constant $\Lambda_0$ on the right-hand side of Eq.~(\ref{ArgPosMS}) and $V_{\rm ns}$ would still  be unbounded from below if there were negative  eigenvalues of $M_S^2$.

Since in the RSB mechanism all $m_s^2$, as well as all $m_f^2$ and $m_v^2$, are non-negative the effective potential is real. This is not the case in theories where symmetry breaking occurs through the standard Higgs mechanism: since the tree-level potential  is not always convex some of the $m_s^2$ are necessarily negative and the effective potential acquires an imaginary part. 
This pathology is a manifestation of the breaking of the  (loop) perturbation theory: it occurs because the loop expansion is an expansion around the minima of the scalar action and the regions where the tree-level potential is not convex are too far from those minima. Therefore, the RSB mechanism supports the validity of perturbation theory in the calculation of the effective potential.

  \section{Phase transition}\label{PT}
  
  We are now ready to study the PT associated with a RSB in our general theory~(\ref{eq:Lmatterns}). In  field theory the role of the order parameter can be played by the expectation value $\langle\chi\rangle$, which includes quantum as well as thermal averages.
  
 \subsection{An RSB phase transition is always of first order}\label{1stPT}
 
\begin{figure}[t]
\begin{center}
  \includegraphics[scale=0.5]{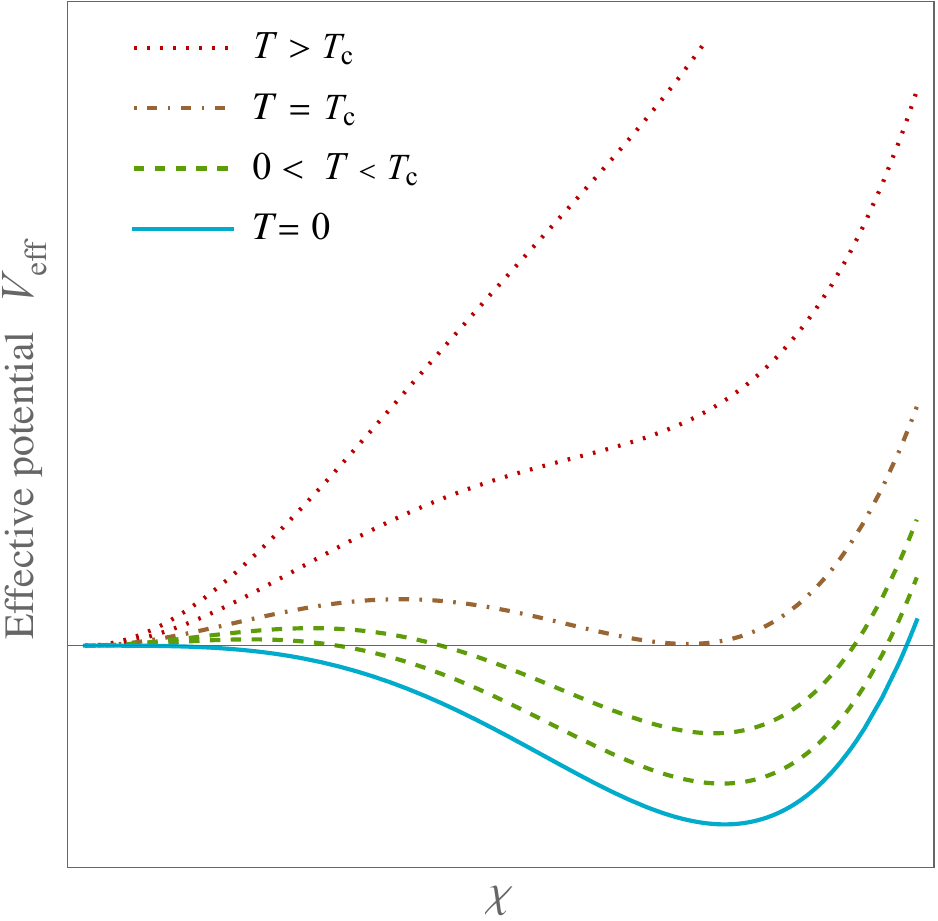}  \hspace{1cm} \includegraphics[scale=0.47]{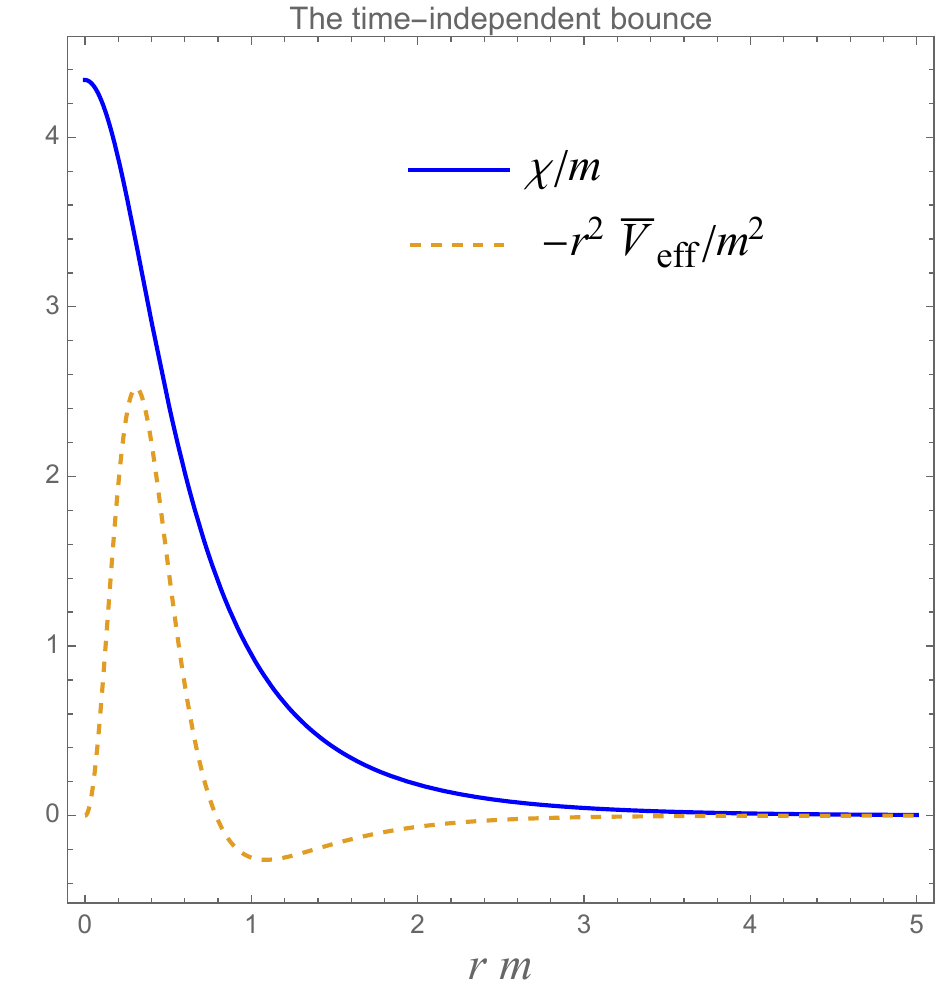}  
    \caption{\em {\bf Left plot:} The temperature-dependent effective potential (including both quantum and thermal contributions) corresponding to a first-order phase transition: the two minima associated with the two phases are separated by a potential barrier; $T_c$ is the critical temperature. {\bf Right plot:} The time-independent bounce and the corresponding integrand function (divided by $8\pi$) appearing in the bounce action, Eq.~(\ref{bounceSd}), for the effective potential $\overline V_{\rm eff}(\chi) = \frac{m^2}{2} \chi^2-\frac{\lambda}{4}\chi^4$ and setting $\lambda=1$.}\label{CWpotf}
  \end{center}
\end{figure}

 The PT associated with a radiative symmetry breaking is always of first order, namely of the type illustrated in the left plot of Fig.~\ref{CWpotf}, as we now show. First, recall that the definition of RSB implicitly assumes the validity of the perturbative expansion. So in this section we start from this assumption. In Sec.~\ref{supercool}, however, it will be shown that supercooling (together with, of course, the assumption of small-enough couplings)  is sufficient to establish the validity of the one-loop approximation to show that, for the relevant temperatures, there is always a barrier separating the two configurations $\chi=0$ and $\chi=\chi_0$.

Note that the first three derivatives of the quantum part of the effective potential, $V_q(\chi)$ in~(\ref{CWpot}), vanishes at the origin, $\chi=0$. On the other hand,  $J_B(x)$ and $J_F(x)$ feature in their small-$x$ expansion a term linear in $x$ with a coefficient that is positive in $J_B$ and negative in $J_F$ (see Eqs.~(\ref{JBdef}) and~(\ref{JFdef})).
Since $J_B$ and $J_F$ appear in the effective potential in the way described by Eq.~(\ref{VeffSumm}), this implies that the effective potential has a minimum at the origin, $\chi=0$, at any non-vanishing temperature.  In other words thermal correction renders $\chi=0$ at least metastable. Going to small enough temperatures the absolute minimum should be approximately the $T=0$ one given by the potential $V_q$, but since there is always a positive quadratic term thanks to the finite-temperature contributions the full effective potential  always features a barrier between $\chi=0$ and $\chi=\chi_0$. 
So the PT is always of first order. Note that this reasoning does not assume the presence of a cubic term, $\sim \chi^3$, in the effective potential, which emerges when bosonic fields are coupled to $\chi$ (see Eq.~(\ref{JBdef})), because it also holds when only fermion fields interact with $\chi$.

The absolute minimum of the effective potential is at $\langle \chi\rangle=0$ for $T$ larger than the critical temperature $T_c$, while, for $T<T_c$, is at a non-vanishing temperature-dependent value. In the latter case the decay rate per unit of spacetime volume, $\Gamma$, of the false vacuum $\langle\chi\rangle=0$ into the true vacuum $\langle\chi\rangle\neq 0$ occurs via quantum and thermal tunnelling through the barrier and can be computed with the formalism of~\cite{Coleman:1977py,Callan:1977pt,Linde:1980tt,Linde:1981zj}: 
\be  \Gamma\sim \exp(-S)\, ,\label{Gamma}\ee
where $S$ is the action
\be S=4\pi\int_0^{1/T} dt_E \int_0^\infty dr r^2\left(\frac12 \dot\chi^2+\frac12 \chi'^2+\bar V_{\rm eff}(\chi,T)\right),  \qquad \bar V_{\rm eff}(\chi,T)\equiv V_{\rm eff}(\chi,T)-V_{\rm eff}(0,T)\label{SVeff}\ee
evaluated at the bounce, which is  the solution of the  differential problem~\cite{Salvio:2016mvj} 
\bea && \qquad \ddot\chi+\chi''+\frac{2}{r}\chi'= \frac{d\bar V_{\rm eff}}{d\chi}, \label{bounceProbE} \\ \dot\chi(r,0)=0, &&\,\dot\chi(r,\pm 1/(2T))=0, \quad  \chi'(0,t_E)=0, \quad \lim_{r\to \infty}\chi(r,t_E) = 0. \label{bounceProb}\eea
Here a dot denotes a derivative with respect to the Euclidean time $t_E$ and a prime denotes a derivative with respect to the spatial radius $r\equiv \sqrt{\vec{x}^{\,2}}$. Indeed, the theory at finite $T$ can be formulated as a theory at imaginary time with time period $1/T$ and the boundary conditions in~(\ref{bounceProb}) impose this periodicity. A particular solution of~(\ref{bounceProbE})-(\ref{bounceProb}) is the time-independent bounce, 
\be \chi''+\frac{2}{r}\chi'= \frac{d\bar V_{\rm eff}}{d\chi}, \qquad \chi'(0)=0, \quad \lim_{r\to \infty}\chi(r) = 0, \label{bounceProb3}\ee
for which
\be S =\frac{S_3}{T}, \qquad  \quad S_3 \equiv 4\pi \int_0^\infty dr \, r^2\left(\frac12 \chi'^2+\bar V_{\rm eff}(\chi,T)\right). \label{S3f}\ee
But generically there could also be time-dependent solutions. When the time-independent bounce dominates the decay rate~\cite{Linde:1980tt,Linde:1981zj}
\be  \Gamma\approx T^4\left(\frac{S_3}{2\pi T}\right)^{3/2}\exp(-S_3/T). \label{Gamma3}\ee
Also note that $S_3$ evaluated at the time-independent bounce can be simplified through the  arguments of~\cite{Coleman:1977th} to obtain 
\be S_3 =  -8\pi \int_0^\infty dr \, r^{2} \bar V_{\rm eff}(\chi,T). \label{bounceSd}\ee
Using the expression above instead of the one in~(\ref{S3f}) makes numerical calculations easier  because the derivatives of $\chi$ do not appear in  the action.
In general bounce solutions describe the production of bubbles of the true vacuum inside a background of false vacuum.

 \subsection{Supercooling and tunneling}\label{supercool}
 
 As long as perturbation theory holds, in a generic theory with RSB, Eq.~(\ref{eq:Lmatterns}),  when $T$ goes below $T_c$ the scalar field $\chi$ is trapped in the false vacuum $\langle\chi\rangle=0$ until  $T$ is much below $T_c$, in other words the universe features a phase of supercooling~\cite{Witten:1980ez}. To understand why this is always the case, note that if the theory is scale invariant  $\Gamma$ must scale as $T^4$ and, therefore, the smaller $T$,  the smaller $\Gamma$. At quantum level, however, scale invariance is broken by perturbative loop corrections, which introduce another dependence of $T$ in the bounce action. This dependence, however, is logarithmic and can become large only when $T$ is very small compared to the other scale of the problem, $\chi_0$.

We also note that, as a consequence of supercooling, the derivative loop corrections to the effective action can be neglected: at $T=0$ the quantum effective potential is very shallow because it is only due to perturbatively small loop corrections. So higher-derivative corrections are very small. Also two-derivative and potential loop corrections are suppressed by loop factors. At $T\neq0$ supercooling implies that for the relevant temperatures the thermal corrections are very small and so $V_{\rm eff}$ is still  very shallow and derivative corrections are very small.

Moreover, supercooling tells us that we are far from the high-temperature regime for which the perturbative expansion is known to break down~\cite{Weinberg:1974hy,Linde:1980ts,Linde:1978px}, so a one-loop computation should be a good approximation. In particular, it should be noted that the fields that participate in the transition (that directly interact with the flat-direction field $\chi$) receive, thanks to supercooling, a zero-temperature mass that is much larger than the thermal mass and the infrared problem discussed in~\cite{Linde:1980ts,Linde:1978px} is avoided. Therefore, we see that the assumption of supercooling also allows us to be confident about the validity of the one-loop approximation.

Remarkably, if supercooling is strong enough in a generic theory of the form~(\ref{eq:Lmatterns}), to good accuracy\footnote{The accuracy of the approximation will be analyzed in Sec.~\ref{impro}.}, the full effective action for relevant values of $\chi$ can be described by three and only three parameters: $\chi_0$, $\bar\beta$ and a real and non-negative quantity $g$ defined as 
follows:
\be g^2 \chi^2\equiv \sum_b n_bm_b^2(\chi)+\sum_f m^2_f(\chi). 
\label{M2g2def}\ee
In other words $g^2\chi^2$ is the sum of all bosonic squared masses plus the sum of all Weyl-spinor squared masses\footnote{Note that $g^2 \chi^2$ defined in~(\ref{M2g2def}) does not coincide with the supertrace of the squared-mass matrix because the fermion masses contribute positively  in~(\ref{M2g2def}).}.
All $m_b^2$ and $m_f^2$ are real, non-negative  and proportional to $\chi^2$, so $g^2$ is real, non-negative and independent of $\chi$.
Note that $g$ plays the role of a ``collective coupling" of $\chi$ with all fields of the theory.

We now show the remarkable property mentioned above. First note that the dominant contributions to the bounce action $S$ are those from field values around the barrier. 
 Therefore, we first need to estimate the barrier size, which we can define as the field value $\chi_b$ at which $V_{\rm eff}$ equals its value at the false vacuum $\chi=0$:
\be \bar V_{\rm eff}(\chi_b,T)  = 0, \ee 
where $\bar V_{\rm eff}$ has been defined in Eq.~(\ref{SVeff}).
Since $\bar V_{\rm eff}$ depends on $T$, the field value $\chi_b$ will be a function of $T$ too. Now let us write the logarithmic term  in the quantum contribution $V_q$ to $V_{\rm eff}$, Eqs.~(\ref{CWpot}) and~(\ref{VeffSumm}), as follows
\be \log\frac{\chi_b}{\chi_0} -\frac14= \log\frac{\chi_b}{T}-\frac14+\log\frac{T}{\chi_0}. \label{logSplit}\ee 
In the presence of supercooling, $T\ll\chi_0$,
 we expect that neglecting the first two terms in the right-hand-side of Eq.~(\ref{logSplit}) is a good approximation because, unlike $\chi_0$, the field value $\chi_b$ clearly becomes small when $T\ll \chi_0$,
\be \log\frac{\chi_b}{\chi_0} -\frac14\approx \log\frac{T}{\chi_0}. \label{logApp}\ee 
If so, using the expression of the effective potential in~(\ref{CWpot}) and~(\ref{VeffSumm}), one finds 
\be \frac{\chi_b^4}{T^4} \approx \frac2{\pi^2}\frac{J_T(\chi_b^2/T^2)-J_T(0)}{\bar\beta\log\frac{\chi_0}{T}},  \label{chibT}\ee
where 
\be J_T(\chi^2/T^2) \equiv \sum_b n_b J_B(m_b^2(\chi)/T^2)-2\sum_f J_F(m_f^2(\chi)/T^2).\ee
The expression in~(\ref{chibT}) tells us that supercooling, $T\ll\chi_0$, suppresses the ratio $\chi_b(T)/T$, but only logarithmically:
\be \frac{\chi_b^4}{T^4} \approx \frac2{\pi^2}\frac{J_T(\chi_b^2/T^2)-J_T(0)}{\bar\beta\log\frac{\chi_0}{T}} \approx \frac2{\pi^2}\frac{J_T'(0)}{\bar\beta\log\frac{\chi_0}{T}}  \frac{\chi_b^2}{T^2} = \frac{g^2}{6\bar\beta \log\frac{\chi_0}{T}}\frac{\chi_b^2}{T^2} \implies \frac{\chi_b^2}{T^2}  \approx \frac{g^2}{6\bar\beta \log\frac{\chi_0}{T}}. \label{chibTorder} \ee
 So the approximation in~(\ref{logApp}) is indeed valid.
Looking now at the effective potential in~(\ref{VeffSumm}) we see that if the quantity $\epsilon$ defined by
\be \epsilon\equiv  \frac{g^4}{6\bar\beta \log\frac{\chi_0}{T}}
 \label{CondConv}\ee
is small 
 we can approximate
\bea J_B(x) &\approx& J_B(0)+\frac{\pi^2}{12} x ,\label{JBapp}  \\
J_F(x) &\approx&J_F(0)-\frac{\pi^2}{24}x, \label{JFapp}\eea
where~(\ref{JBdef}) and~(\ref{JFdef}) have been used. In~(\ref{CondConv}) an extra factor $g^2$ has been inserted compared to~(\ref{chibTorder}) because $\chi^2/T^2$ appear in the thermal functions multiplied by some coupling constant.
Note that the approximations in~(\ref{JBapp}) and~(\ref{JFapp}) are not valid for all values of $\chi$, including $\chi_0$, because of supercooling, $T\ll \chi_0$. But, as we have just shown, they are valid for the field values that are important in the bounce action if a large-enough supercooling occurs ($\epsilon$ small). This is because in this case $g^2 \chi_b^2/T^2$ is small (see the last equation in~(\ref{chibTorder})).
Now, using the approximations in~(\ref{logApp}),~(\ref{JBapp}) and~(\ref{JFapp}), the bounce action can be computed with the effective potential given by
\be \bar V_{\rm eff}(\chi,T) \approx \frac{m^2(T)}{2} \chi^2-\frac{\lambda(T)}{4} \chi^4 \label{barVapp}\ee
where $m$ and $\lambda$ are real and positive functions of $T$ defined by
\be m^2(T) \equiv \frac{g^2 T^2}{12}, \qquad \lambda(T) \equiv \bar\beta \log\frac{\chi_0}{T} \label{mlambdaDef}\ee
and $g^2$ is the collective coupling  defined in~(\ref{M2g2def}).


We can now see that the tunneling process is dominated by the time-independent bounce, which satisfies~(\ref{bounceProb3}). The expression of $\bar V_{\rm eff}$ in~(\ref{barVapp}), together with the form of the bounce problem in~(\ref{bounceProbE})-(\ref{bounceProb}), tells us that the characteristic bounce size $R_b$ is of order $R_b\sim 1/m(T)\gtrsim1/T$, where in the second estimate we have used the perturbativity condition that $g$ is not too large. This result tells us that  the bounce solutions  are approximately time-independent (see~\cite{Linde:1980tt,Linde:1981zj}). Moreover, for a time-dependent bounce the Euclidean action $S$ turns out to be larger than the one  of the time-independent bounce for all values of $\lambda$ and $g$ (at least in the perturbative domain)~\cite{Salvio:2016mvj}.  This means that the tunneling process is dominated by the time-independent bounce. 

At this point we can also note that the gravitational corrections to the false vacuum decay are amply negligible whenever the symmetry breaking scale $\chi_0$ is small compared to the Planck mass $M_P$, which is, of course, the most interesting case from the  phenomenological point of view. This is because, as we have seen, the temperature and  the typical scales of the bounce are always much below $\chi_0$ and the gravitational corrections are, therefore, suppressed by factors much smaller than 
 $\chi_0^2/\bp^2$~\cite{Salvio:2016mvj}, where $\bp$ is the reduced Planck mass that is defined in terms of the Planck mass $M_P$ by $\bp \equiv M_P/\sqrt{8\pi}$.

Note now that the bounce action $S_3$ computed with our effective potential in~(\ref{barVapp}) is a function of $m$ and $\lambda$ only,  $S_3=S_3(m,\lambda)$. If we rescale $\chi\to\chi/\sqrt{\lambda}$  one finds $S_3(m,\lambda)=S_3(m,1)/\lambda$. Also, using dimensional analysis 
\be S_3=c_3 \frac{m}{\lambda}, \label{S3c3}\ee where $c_3$ is a dimensionless number: computing explicitly the bounce for $\lambda=1$ (see the right plot of Fig.~\ref{CWpotf}) and its action through Eq.~(\ref{bounceSd}) we find 
\be c_3=-\frac{8\pi}{m}
 \int_0^\infty dr \, r^2 \left( \frac{m^2}{2} \chi^2-\frac{1}{4} \chi^4\right)
=18.8973... \label{c3value}\ee
 (see also~\cite{Brezin-Parisi,Arnold:1991cv} for previous calculations). This quite large value of $c_3$ is due to the geometrical factor of $4\pi$ overall. The right plot of Fig.~\ref{CWpotf} shows that for the values of $r$ that give the largest contribution to the bounce action the quartic term is significantly bigger than the quadratic one. 

During supercooling the energy density  is dominated by the vacuum energy of $\chi$ and the universe grows exponentially with Hubble rate $H_I = \sqrt{\bar\beta} \chi_0^2/(4\sqrt{3}\bp)$. The bubbles created are   diluted by the expansion of the universe and they cannot collide until $T$ reaches the nucleation temperature $T_n$, which corresponds to $\Gamma/H_I^4 \sim 1$ or, equivalently, using the fact that the decay is dominated by the time-independent bounce,
\be  \frac{S_3}{T_n}-\frac32 \log \left(\frac{S_3/T_n}{2\pi}\right) \approx 4 \log \left(\frac{T_n}{H_I}\right),\label{TnEq0}\ee
By using the expression of $S_3$ in~(\ref{S3c3}) and the definitions in~(\ref{mlambdaDef}) one finds
\be \frac{c_3 g}{\sqrt{12}\bar\beta\log\frac{\chi_0}{T_n}} -\frac32\log\left(\frac{c_3 g}{2\sqrt{12}\pi\bar\beta\log\frac{\chi_0}{T_n}}\right) \approx 4\log\left(\frac{4\sqrt{3} \bp T_n}{\sqrt{\bar\beta}\,\chi_0^2}\right)\ee
or, equivalently, the approximate equation in $X$
\be \frac32 X \log X \approx c X - 4 X^2 - a \label{TnEq}\ee 
having  defined
\be X\equiv \log\frac{\chi_0}{T_n} \label{Xdef}\ee
and
\be a\equiv \frac{c_3g}{\sqrt{12}\bar\beta}, \quad c\equiv 4\log\frac{4\sqrt{3}\bp}{\sqrt{\bar\beta}\,\chi_0}+\frac32\log\frac{a}{2\pi}. \label{caDef}\ee
Recall that $g$ is never negative, the number $c_3$ has the positive value in~(\ref{c3value}) and $\bar\beta>0$, see~(\ref{eq:CWgen}), so $a$ is never negative.

At this point it is important to note that quantities of order $\log X$ has been considered negligible  compared to terms of order $X$ in the approximations around Eqs.~(\ref{logSplit})-(\ref{chibTorder}).  To be consistent with these approximations we drop the term  $\frac32 X\log X$ in~(\ref{TnEq}):
\be c X - 4 X^2 - a \approx 0. \label{TnEq2}\ee
Here we are interested in the solution of Eq.~(\ref{TnEq2}) with the smaller $X$, which corresponds to $\Gamma$ reaching $H_I^4$ from below. This solution  gives
\be T_n\approx \chi_0\exp\left(\frac{\sqrt{c^2-16a}-c}8\right). \label{appTn}\ee
Note that in the decoupling limit ($g\to0$, $\bar\beta/g\to0$ and $\chi_0$ fixed) $\sqrt{a}\to\infty$ faster than $c\sim \log a$ 
and there is no solution for $T_n$. This occurs because in this limit $\Gamma\to 0$ and so can never be of order $H_I^4$. As a result, the existence of a solution of Eq.~(\ref{TnEq2}), which determines $T_n$, requires a minimum value of the collective coupling $g$, such that $c^2\geq 16a$; however, the smaller $g$ (at $\bar\beta/g^4$ fixed) the larger $\log(\chi_0/T_n)$  and the universe supercools more, at least for realistic and perturbative values of the parameters.
Moreover, in general supercooling is also enhanced by increasing $\chi_0$ at $g$ and $\bar\beta$ fixed, although this effect is very mild as $c$ depends on $\chi_0$ only logarithmically.
In Sec.~\ref{impro} we will discuss the accuracy of the approximations performed in this section and explain how to improve them.

One might wonder whether the effect of the spacetime curvature due to $H_I\neq 0$ can alter the decay rate. In standard Einstein gravity, this may happen if $T_n$ is so small to be comparable with $H_I$. We checked that, whenever a solution for $T_n$ exists, this never happens, at least for realistic and perturbative values of the parameters. On the other hand, if a solution for $T_n$ does not exist  the effect of the spacetime curvature, as well as quantum fluctuations, can eventually become important in the decay rate~\cite{Kearney:2015vba,Joti:2017fwe,Markkanen:2018pdo,DelleRose:2019pgi}.

In general, the strength of the PT is measured by the parameter $\alpha$ defined as the ratio between 
\be \rho(T_n) \equiv \left[\frac{T}{4} \frac{d}{dT} \bar V_{\rm eff}(\langle\chi\rangle,T)-\bar V_{\rm eff}(\langle\chi\rangle,T)\right]_{T=T_n} \label{rhoDef} \ee
and the energy density of the thermal plasma (see~\cite{Caprini:2019egz,Ellis:2019oqb} for more details). In~(\ref{rhoDef}) $\langle\chi\rangle$ is the point of absolute minimum of the effective potential, which at $T=T_n$ is not zero.  By definition $\alpha$ is
\be \alpha \equiv \frac{30 \rho(T_n)}{\pi^2 g_*(T_n)T_n^4}, \ee
where $g_*(T)$ is the effective number of relativistic species at temperature $T$.
In the presence of supercooling we then typically have 
\be \rho(T_n) \approx \left[-\bar V_{\rm eff}(\langle\chi\rangle,T)\right]_{T=T_n}.  \ee
Moreover,
\be \alpha \gg 1,\ee
because the energy density is dominated by the vacuum energy of $\chi$, as we have seen.
 So in the RSB scenario one has  a very strong PT.

 \section{Gravitational Waves}\label{Gravitational Waves}

   In the RSB scenario the dominant source of GWs are bubble collisions that take place in the vacuum. This is because the energy density of the space 
  where the bubbles move are dominated by the vacuum energy density associated with $\chi$, which leads to an exponential growth of the corresponding cosmological scale factor as we have seen. This inflationary behavior as usual dilutes preexisting matter and radiation and, therefore, we neglect the GW production due to turbulence and sound waves  in the cosmic fluid~\cite{Caprini:2015zlo,Maggiore:2018sht,Lewicki:2022pdb}.

 An important parameter to analyse the spectrum of the GWs  is the inverse duration $\beta$ of the PT that, in models with supercooling, is given by~\cite{Caprini:2015zlo,Caprini:2018mtu,vonHarling:2019gme} 
  \be\beta = \left[\frac1{\Gamma}\frac{d\Gamma}{dt}\right]_{t_n},\ee 
  where   $t_n$ is the value of the time $t$ when $T=T_n$. Using then $dt = -dT/(TH(T)) $, where $H(T)$ is the Hubble rate corresponding to the temperature $T$, 
  and the fact 
that the tunneling process is dominated by the time-independent bounce,
\be \beta \approx H_n\left[T\frac{d}{dT}(S_3/T)-4-\frac{3}{2}T\frac{d}{dT}\log(S_3/T)\right]_{T=T_n}, \label{betaH1}\ee 
where $H_n\approx H_I$ is the Hubble rate  when $T=T_n$. 

 Now, if supercooling is large enough that one can use the expression of $S_3$ in~(\ref{S3c3}),
 one obtains 
\be \frac{\beta}{H_n} \approx \frac{a}{\log^2(\chi_0/T_n)} -4 -\frac32\frac1{\log(\chi_0/T_n)}, \label{betaH2s} \ee
where $a$ is defined in terms of $\bar\beta$ and $g$ in~(\ref{caDef}).
The last term in the expression above, which corresponds to the last term in~(\ref{betaH1}), can be neglected as $\chi_0/T_n$ is very large. However, the first term $a/\log^2(\chi_0/T_n)$ generically cannot be neglected because the coefficient $a$ is typically very large as $\bar\beta$ is loop suppressed and $c_3$ is larger than 10, see~(\ref{c3value}). So 
\be \frac{\beta}{H_n} \approx \frac{a}{\log^2(\chi_0/T_n)} -4. \label{betaH2} \ee
  Eq.~(\ref{betaH2}) then explicitly indicates that the PT lasts more when $T_n/\chi_0$ is smaller. 
  Inserting the expression of $T_n$ given in~(\ref{appTn}) into~(\ref{betaH2}),
  one obtains $\beta/H_n$ in terms  of $\chi_0$, $\bar\beta$  and $g$.

  From~\cite{Caprini:2015zlo} we find the following GW spectrum 
due to vacuum bubble collisions\footnote{The spectral density $\Omega_{\rm GW}$ is defined as usual by
\be \Omega_{\rm GW}(f) \equiv \frac{f}{\rho_{\rm cr}}\frac{d\rho_{\rm GW}}{df},\ee 
where $\rho_{\rm cr}\equiv	3H_0^2\bp^2$  is the critical energy density, $H_0$ is the present value of the Hubble rate and $\rho_{\rm GW}$ is the energy density carried by the stochastic background.}
 (valid in the presence of supercooling and $\alpha\gg 1$)
\be \label{eq:gw_col} h^2 \Omega_{\rm GW}(f) \approx 1.29 
\tt 10^{-6}\left(\frac{H_r}{\beta}\right)^2\left(\frac{100}{g_*(T_r)}\right)^{1/3}\frac{3.8(f/f_{\rm peak})^{2.8}}{1+2.8(f/f_{\rm peak})^{3.8}},\ee 
where $T_r$ is the reheating temperature after supercooling, $H_r$ is the corresponding Hubble rate  and $f_{\rm peak}$ is the red-shifted frequency peak today, which is given by~\cite{Caprini:2015zlo} 
\be f_{\rm peak} \approx 3.79\, \frac{\beta}{H_r}\left( \frac{g_*(T_r)}{100}\right)^{1/6}\frac{T_r}{10^{8}{\rm GeV}}  \, {\rm Hz} . \ee
Ref.~\cite{Caprini:2015zlo} used, among other things, the results of~\cite{Huber:2008hg}  based on the envelope approximation. This is an  approximation where all the energy is assumed to be stored in the bubble walls, which are taken to be thin, and, as the bubbles collide, one uses as a source for GW production the energy-momentum tensor of the uncollided part of the bubble walls. In our situation this is expected to capture the dominant source  of GWs~\cite{Freese:2022qrl} because, during the exponential growth of the universe, the bubbles expand considerably and in this process the energy gained in the transition  from the false to the true vacuum is transferred to the bubble walls, which, at the same time, become thinner
\footnote{See, however, the recent works~\cite{Jinno:2017fby,Konstandin:2017sat,Lewicki:2019gmv,Lewicki:2020jiv,Lewicki:2020azd} that improved the calculation of $\Omega_{\rm GW}$ and can be relevant in the general case.}. 

Eq.~(\ref{eq:gw_col}) shows that the GW spectrum is larger for smaller values of $\beta$. So an approximate scale invariance, which, as we have seen, can suppress 
 $\beta$, can also lead to strong GW signals;  the larger the supercooling is the stronger the GW signals are.
  For sufficiently fast reheating, which might occur e.g. thanks to the Higgs portal coupling in~(\ref{portal}),
\be H_r\approx H_n \approx H_I, \quad \mbox{and} \qquad T_r^4 \approx \frac{15 \bar\beta \chi_0^4}{8\pi^2 g_*(T_r)}.\label{TRHmax}\ee But otherwise $H_r$ and $T_r$ can depend on the details of the  specific model.
  
Note that any cosmic source of GW background acts as an extra radiation component and is highly constrained by big-bang nucleosynthesis (BBN) measurements of primordial elements~\cite{Boyle:2007zx,Stewart:2007fu,Kohri:2018awv}.
 The measurement of the effective number of neutrino species $N_{\rm eff}$ and the observational abundance of deuterium and helium, leads to  
 the following bound~\cite{Maggiore:2018sht}:
 \be \int_{f_{\rm BBN}}^{f_{\rm UV}} \frac{df}{f} h^2 \Omega_{\rm GW}(f) <1.3 \tt 10^{-6}  \, \frac{N_{\rm eff}-3.046}{0.234},\label{BBNbound}\ee
where $f_{\rm BBN}\sim 10^{-11} $Hz (see e.g.~\cite{Boyle:2007zx,Maggiore:2018sht}) and $f_{\rm UV}$ is some UV cutoff, which  can be conservatively taken to be around $\bp$. In the right-hand-side of~(\ref{BBNbound}) we can use the reference value $N_{\rm eff} =2.99+0.17=3.16$, which corresponds to the most precise experimental upper bound on $N_{\rm eff}$ published by the Planck collaboration in 2018~\cite{Planck:2018vyg}.

\begin{figure}[t!]
\begin{center}
  \includegraphics[scale=0.60]{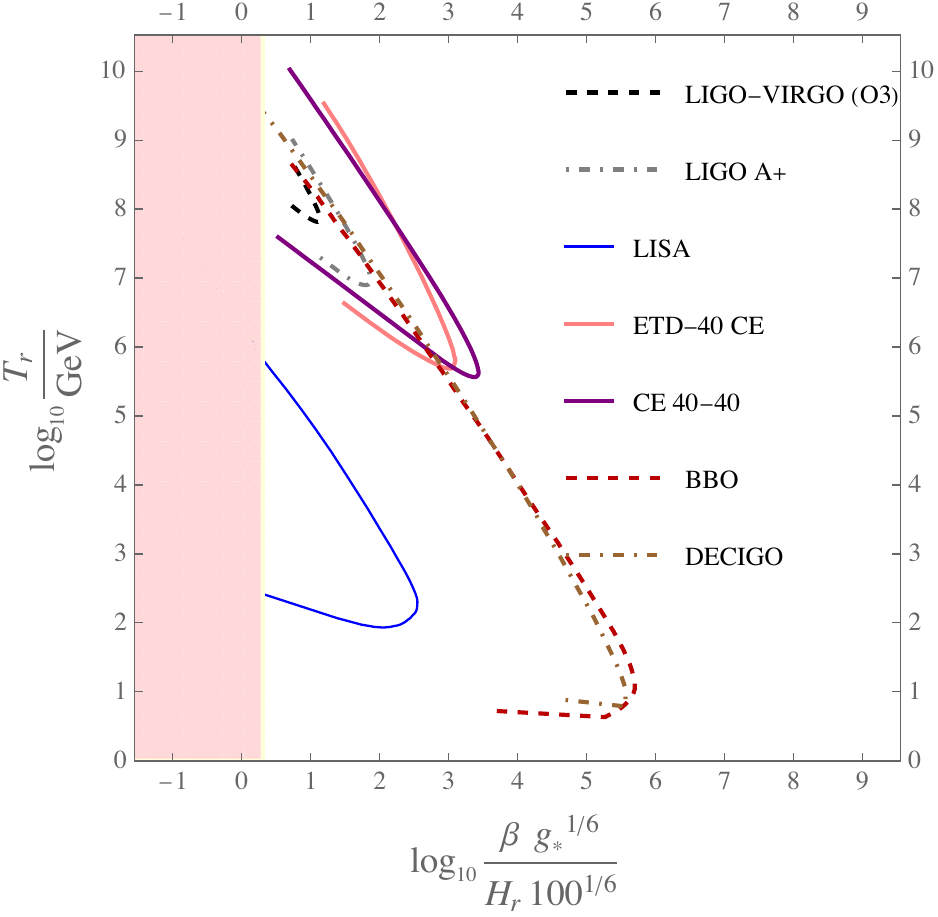} \qquad \quad  \includegraphics[scale=0.51]{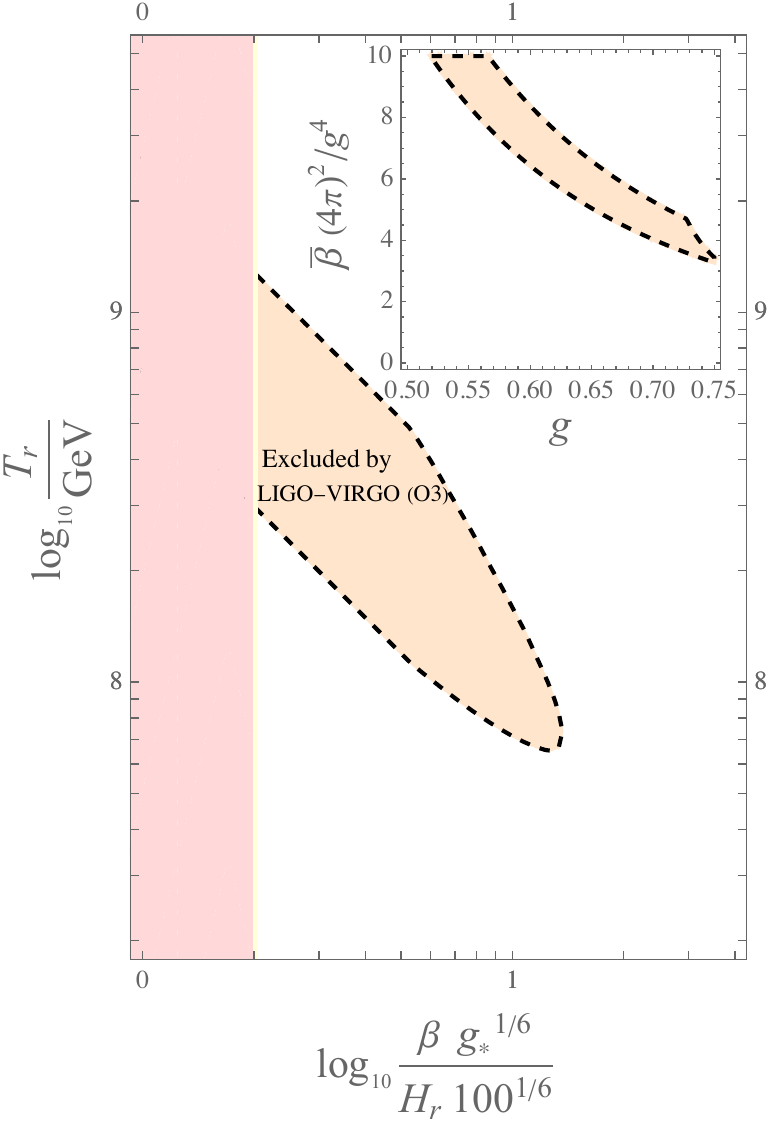}
      \caption{\em Regions where $\Omega_{\rm GW}(f_{\rm peak})$ is above the sensitivities of several current and proposed GW detectors. The shaded pink region to the left of the vertical straight line is the one forbidden by the BBN bound in~(\ref{BBNbound}). 
      {\bf Left plot:} For each curve, the region  is the one obtained drawing semi-straight lines with derivative equal to -1 starting from all points of the curve and going to the left. {\bf Right plot}: a zoom of the region corresponding to the LIGO-VIRGO third observing run (O3), which has therefore been excluded; the inset in the right plot shows the corresponding excluded region in the $\{g,\bar\beta\}$ space in the case of fast reheating (fixing $g_*(T_r)=110$ and $\chi_0 = 2\times 10^9$~GeV)} \label{betaVT}
  \end{center}
\end{figure}

Fig.~\ref{betaVT} shows the regions where $\Omega_{\rm GW}(f_{\rm peak})$
 is above the sensitivities of several current and proposed GW detectors: Advanced LIGO's and Advanced Virgo's third observing (O3) run (for power law-spectra), a possible upgrade of the current Advanced LIGO facilities (LIGO A+)~\cite{KAGRA:2021kbb,LIGOScientific:2023vdi}, LISA (power law sensitivity)~\cite{Babak:2021mhe}, an ET array combined with a CE in the US (ETD-40 CE) and two CE detectors in the US with arm lengths  of 40 km~\cite{Evans:2021gyd}, BBO and DECIGO (for power law spectra)~\cite{Thrane:2013oya,Dev:2019njv}. In Fig.~\ref{betaVT} it is also shown the bound in~(\ref{BBNbound}). 
 
\begin{figure}[t]
\begin{center}
 \vspace{1cm}
  \includegraphics[scale=0.52]{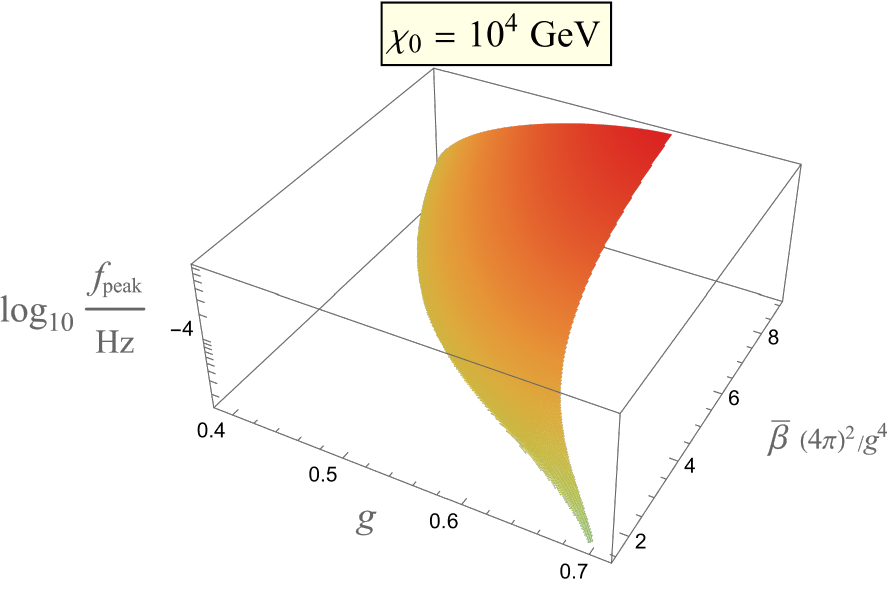}  \hspace{1cm}\includegraphics[scale=0.54]{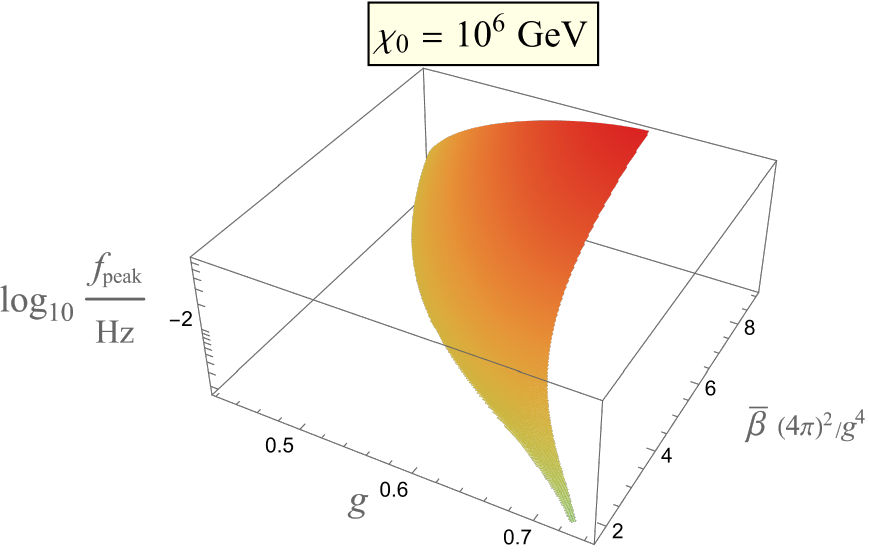} \\
  \vspace{1cm}
  \includegraphics[scale=0.52]{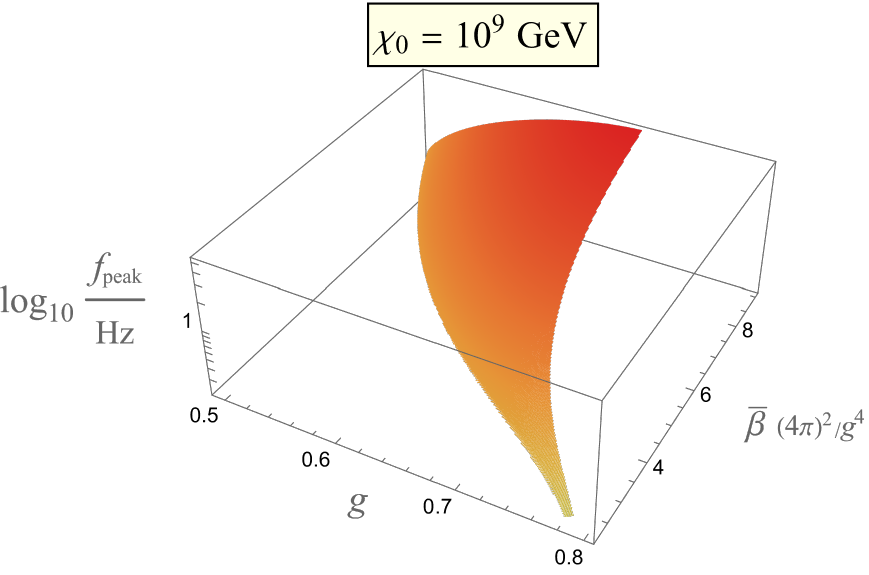}  \hspace{1cm}\includegraphics[scale=0.54]{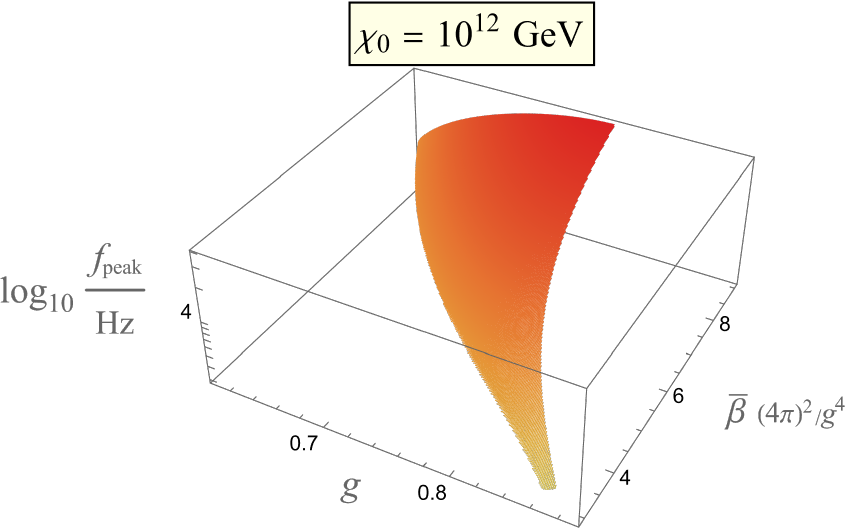}   
      \caption{\em The peak frequency as a function of $g$ and $\bar\beta$ in the case of fast reheating and fixing $g_*(T_r)=110$.}\label{fp}
  \end{center}
\end{figure}

 The region corresponding to the LIGO-VIRGO O3 is ruled out (a zoom of this region, computed for power-law spectra~\cite{KAGRA:2021kbb}, is given in the right plot of Fig.~\ref{betaVT}), while the other regions, which is most of the parameter space, are still allowed as the corresponding observations have not yet been performed. The inset of the right plot of Fig.~\ref{betaVT} shows the region excluded by the LIGO-VIRGO O3 in the $\{g,\bar\beta\}$ space in the case of fast reheating and fixing $g_*(T_r)$ and $\chi_0$ to some reference values. The latter parameter has been chosen around $10^9$~GeV because the corresponding $f_{\rm peak}$ is then around the frequency range of LIGO-VIRGO O3~\cite{KAGRA:2021kbb} (see Fig.~\ref{fp}). The dependence on $g_*(T_r)$, on the other hand, is very weak. A $\chi_0$ around  $10^9$~GeV is relevant e.g. for axion models. We checked that for all values of $g$ and $\bar\beta$ in the inset of Fig.~\ref{betaVT} the parameter $\epsilon$ in~(\ref{CondConv}) is below one. 
 
 In Fig.~\ref{fp} it is also shown $f_{\rm peak}$ for other values of $\chi_0$; also in that figure we considered only values of $g$ and $\bar\beta$ such that $\epsilon$ is below one. 
 
\begin{figure}[t]
\begin{center}
\includegraphics[scale=0.48]{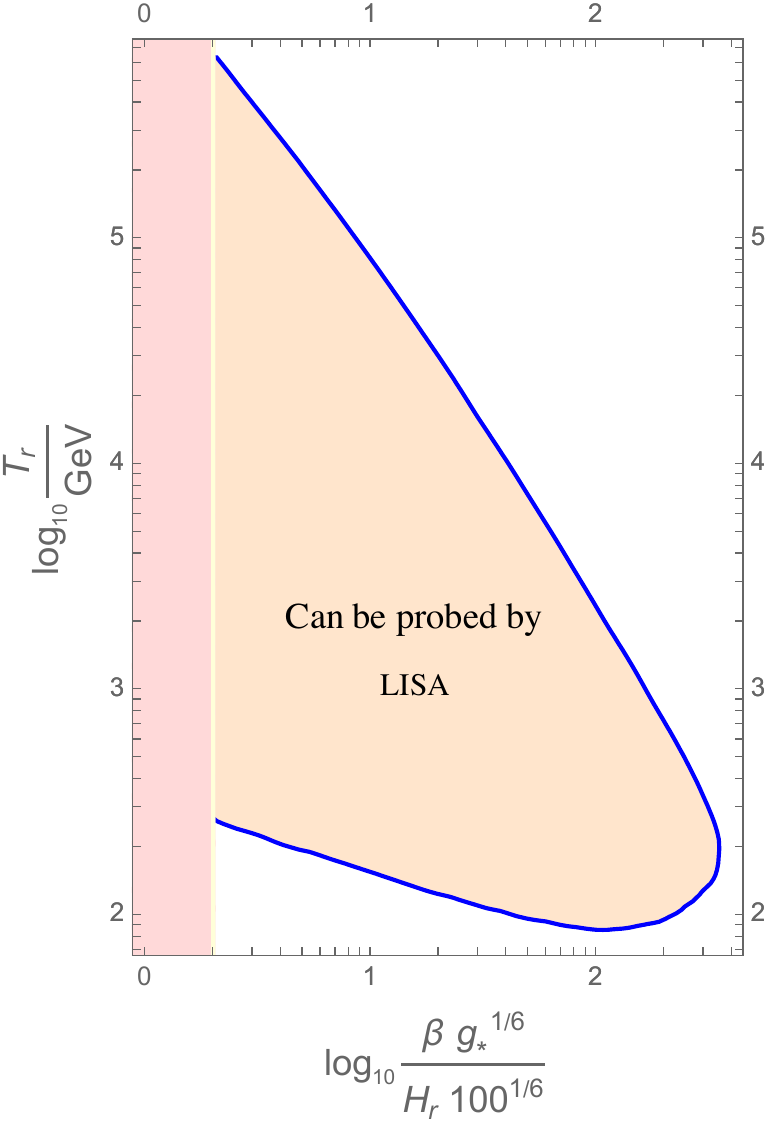}  \hspace{1.8cm}
  \includegraphics[scale=0.48]{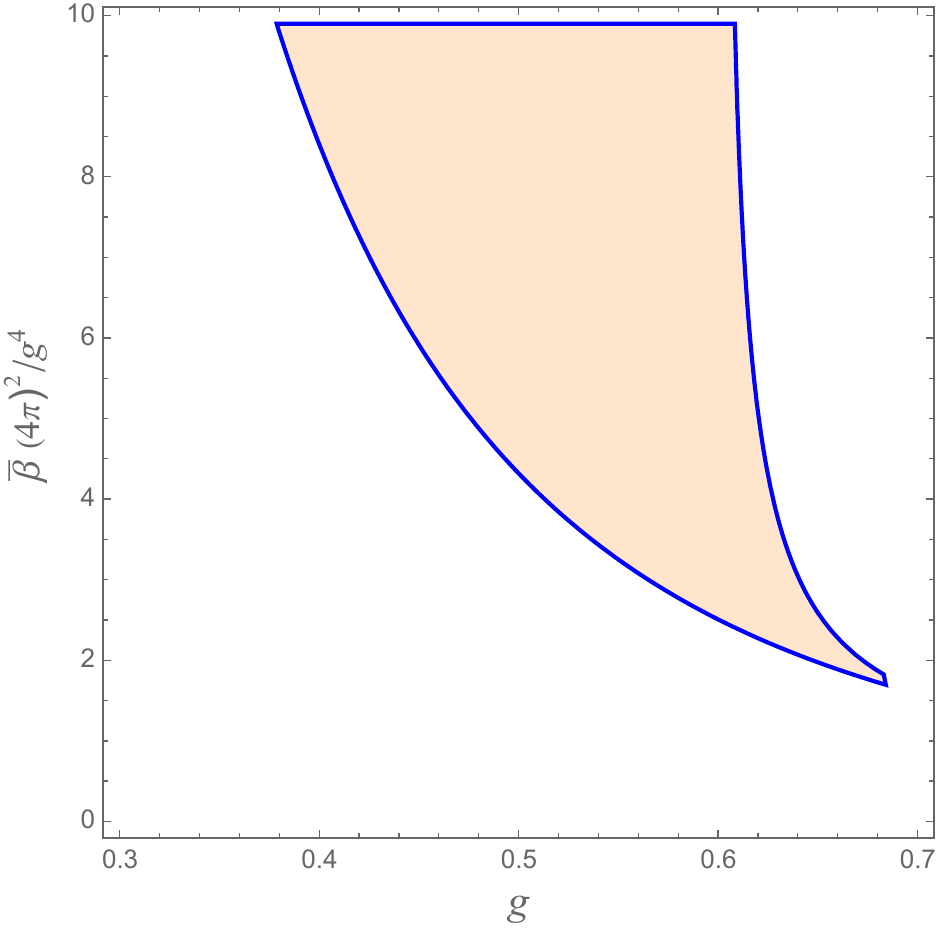}     
      \caption{\em Like in the right plot of Fig.~\ref{betaVT}, but for LISA rather than LIGO-VIRGO O3 and fixing $g_*(T_r)=110$ and $\chi_0 = 10^4$~GeV. }\label{LISA}
  \end{center}
\end{figure}

Using Fig.~\ref{fp} one sees that  when $\chi_0$ is instead closer to the EW scale one will be able to probe this scenario through e.g. LISA, BBO and DECIGO (see Refs.~\cite{Babak:2021mhe,Thrane:2013oya,Dev:2019njv}); higher values of $\chi_0$ can instead be tested by ET and CE (see Ref.~\cite{Evans:2021gyd}). 
 Given the phenomenological relevance of $\chi_0$ close to the EW scale, in Fig.~\ref{LISA} it is shown the region of the parameter space that can be probed by LISA (with the same conventions as in the right plot of Fig.~\ref{betaVT}, but for LISA instead of LIGO-VIRGO O3 and also there we checked that for all values of $g$ and $\bar\beta$ the quantity  $\epsilon$ is below one). A $\chi_0$ around 10 or 100 TeV  is relevant for unified theories such as the Pati-Salam model~\cite{Pati:1974yy}
  or Trinification~\cite{Babu:1985gi}. 
 
\section{Improved approximations}\label{impro}

The approximation performed in~(\ref{JBapp})-(\ref{JFapp}) generically corresponds to neglecting terms of order $\sqrt{\epsilon}$ (where $\epsilon$ is defined in~(\ref{CondConv})). Since we eventually need to set $T=T_n$, see Eq.~(\ref{betaH1}), the approximation in~(\ref{logApp}) and dropping the $X\log X$ term in Eq.~(\ref{TnEq})
for the nucleation temperature $T_n$, on the other hand,  correspond to neglecting terms of relative order $(\log X)/X$, which are smaller than $\sqrt{\epsilon}$ because, since $\epsilon$ is small,  $X\gtrsim g^4/(6\bar\beta) = \epsilon X$, which is large because $\bar\beta$ is loop suppressed\footnote{The approximation in~(\ref{logApp}) consists in neglecting terms of relative order $1/X$ and $\log(\epsilon/g^2)/X$. The order of magnitude of the former is obviously not larger  than $(\log X)/X$, but also that of the  latter is so: apparently it might become larger than $(\log X)/X$ when $g\to 0$, but, as we have seen around~(\ref{appTn}), in this limit there is no solution for $T_n$ and, in any case, $X$ asymptotically goes as an inverse power of $g$ because $\bar\beta \sim g^4/(4\pi)^2$. }. What we are doing here is a small $\epsilon$ expansion (a ``supercool expansion") and what we have done so far is the analysis at leading order (LO), that is modulo terms of relative order $\sqrt{\epsilon}$.
In this section we discuss how to improve the LO result by calculating higher-order corrections in the supercool expansion.

The first step is to calculate the corrections of relative order $\sqrt{\epsilon}$ with respect to the LO approximation, which
 corresponds to working at next-to-leading order  in the supercool expansion (henceforth NLO). This can be done by improving the approximation in~(\ref{JBapp})-(\ref{JFapp}): including the term of order\footnote{This term was overlooked in Ref.~\cite{Witten:1980ez} (see Eq.~(3) there).} $x^{3/2}$ in the expansion of $J_B(x)$ in~(\ref{JBdef}), one obtains
\bea J_B(x) &\approx& J_B(0)+\frac{\pi^2}{12} x-\frac\pi{6} x^{3/2} ,\label{JBappnlo}  \\
J_F(x) &\approx&J_F(0)-\frac{\pi^2}{24}x. \label{JFappnlo}\eea
Since for field values that are relevant for the bounce action $x$ is at most of order $\epsilon$ (see Eqs.~(\ref{chibTorder}) and~(\ref{VeffSumm})), this corresponds to neglecting terms of order not larger than $\epsilon\log \epsilon$ relatively to the LO. So the approximation in~(\ref{logApp}) and dropping terms of order $X\log X$  in the equation for the nucleation temperature are still valid even at NLO.

The effective potential at NLO, therefore, includes a term that is cubic in the field $\chi$ and reads
\be \bar V_{\rm eff}(\chi,T) \approx \frac{m^2(T)}{2} \chi^2-\frac{k(T)}{3}\chi^3-\frac{\lambda(T)}{4} \chi^4 \label{barVnlo}\ee
where $m^2$ and $\lambda$ are defined 
in~(\ref{mlambdaDef}), 
\be k(T)\equiv \frac{\tilde g^3 T}{4\pi}, \ee
and $\tilde g$ is a non-negative real parameter  defined by
\be \tilde g^3\chi^3 \equiv \sum_b n_bm_b^3(\chi). \label{gtdef}\ee
Here the $m_b^3$ are the cube of the background-dependent bosonic squared masses, which are all real, non-negative  and proportional to $\chi$. So $\tilde g^3$ is real, non-negative and independent of $\chi$; it is an extra parameter, which is needed to describe this scenario in a model-independent way at NLO. In general we have
\be \tilde g\leq g. \label{disggt}\ee

 In order to understand why the term cubic in $\chi$ in~(\ref{barVnlo}) can be considered as a small correction in the supercool expansion, one can rescale $\chi\to \chi/\sqrt{\lambda}$ in the bounce action, Eq.~(\ref{SVeff}), to obtain 
\be S = \frac{4\pi}{\lambda}\int_0^{1/T} dt_E \left[ \int_0^\infty dr \, r^2\left(\frac12 \dot\chi^2+\frac12 \chi'^2+\frac{m^2}{2} \chi^2-\frac{1}{4} \chi^4\right) -\frac{k}{3\sqrt{\lambda}} \int_0^\infty dr \, r^2\chi^3 \right].\label{SNLO}\ee 
Since $\lambda=\bar\beta\log(\chi_0/T)$ and eventually we need to set $T=T_n$, we explicitly see that the term proportional to $k$ has relative order at most $\sqrt{\epsilon}$ times a number smaller than one $\approx 1/(\sqrt{2}\pi)$ (where the LO result $S\approx 4\pi gT/(\sqrt{12}\lambda)$ and~(\ref{disggt}) have been used). This small number helps the convergence of the supercool expansion. Working at NLO, we can substitute $\chi$ with the solution of the unperturbed (that is LO) bounce problem both in the second integral in the square bracket (because suppressed by $\sqrt{\epsilon}$) and in the first one (because the first variation of the action around a solution of the field equations vanishes). So one obtains
\be S_3 = \frac{1}{\lambda}\left(c_3 m -\tilde c_3\frac{k}{3\sqrt{\lambda}}\right), \ee
where $c_3$ is given in Eq.~(\ref{c3value}),
\be \tilde c_3 \equiv 4\pi \int_0^\infty  dr \,  r^2\chi_{\rm LO}^3 \ee
and $\chi_{\rm LO}$ is the LO bounce configuration. A numerical calculation then gives $\tilde c_3= 31.6915 ... \,$. Again, like in Eq.~(\ref{c3value}), we find a quite large value because of the  geometrical factor of $4\pi$ overall.

Having obtained the NLO bounce action we can now improve the LO equation in~(\ref{TnEq2}) which gives $T_n$. Using~(\ref{TnEq0}), which holds at all orders in the supercool expansion, one obtains
\be \frac32 X \log X-\frac32 X\log\left(1-\frac{\delta}{a\sqrt{X}} \right)\approx c X - 4 X^2 - a +\frac{\delta}{\sqrt{X}},\label{TnEqNLO0}\ee
where $a$ and $c$ are defined in~(\ref{caDef}) and
\be \delta \equiv \frac{\tilde c_3 \tilde g^3}{12\pi \bar\beta^{3/2}}. \ee
Using~(\ref{disggt}), the quantity $\frac{\delta}{a\sqrt{X}}$ turns out to be at most of order $\sqrt{\epsilon}$. 
So, working at NLO, we can drop both terms on the left-hand side of~(\ref{TnEqNLO0}), which are of relative order $(\log X)/X$ or smaller, and obtain
\be -\frac{\delta}{\sqrt{X}}\approx c X - 4 X^2 - a.\label{TnEqNLO}\ee
Now, to find a solution one can again proceed perturbatively: calling $X_{\rm LO}$ the LO solution, namely
\be X_{\rm LO} = \frac{c-\sqrt{c^2-16a}}8,\ee 
the NLO solution can be obtained by substituting $X$ with $X_{\rm LO}$ only in the left-hand side of Eq.~(\ref{TnEqNLO}) and then solving with respect to $X$. This leads to the following NLO solution for $T_n$:
\be T_n\approx \chi_0\exp\left(\frac{\sqrt{c^2-16(a-\delta/\sqrt{X_{\rm LO}})}-c}8\right). \label{appTnnlo}
\ee

Let us now determine the NLO expression for $\beta/H_n$, Eq.~(\ref{betaH1}). To this purpose it is useful to write 
\be \frac{S_3}{T} = \frac{a}{\log\frac{\chi_0}{T}} \left(1-\frac{\delta/a}{\sqrt{\log\frac{\chi_0}{T}}}\right). \ee 
Inserting now in~(\ref{betaH1}) and dropping terms of relative order smaller than $\epsilon$, which are negligible even at NLO, 
\be \frac{\beta}{H_n} \approx \frac{a}{\log^2(\chi_0/T_n)}\left(1-\frac{3\delta/a}{2\sqrt{\log(\chi_0/T_n)}}\right) -4. \label{betaH2nlo} \ee
Note that, interestingly, the NLO correction reduces $\beta$ and so renders $\Omega_{\rm GW}$ larger, see Eq.~(\ref{eq:gw_col}).

 \begin{figure}[t]
\begin{center}
      \includegraphics[scale=0.5]{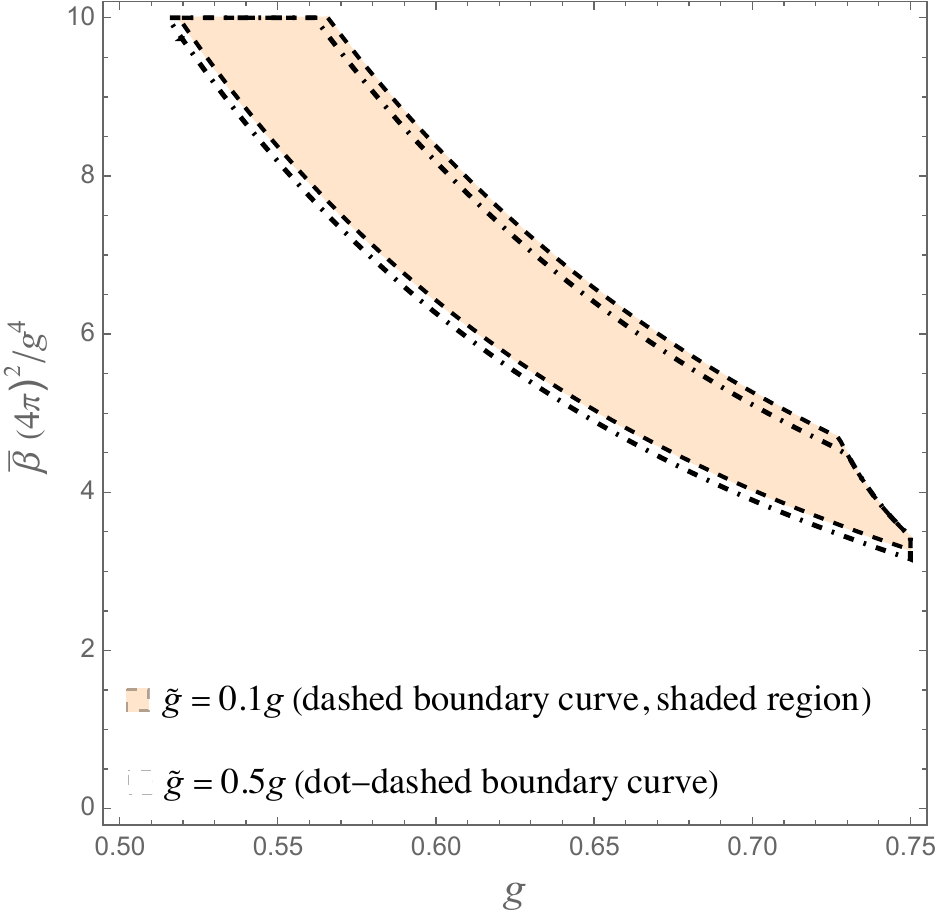}  \hspace{1cm}
  \includegraphics[scale=0.5]{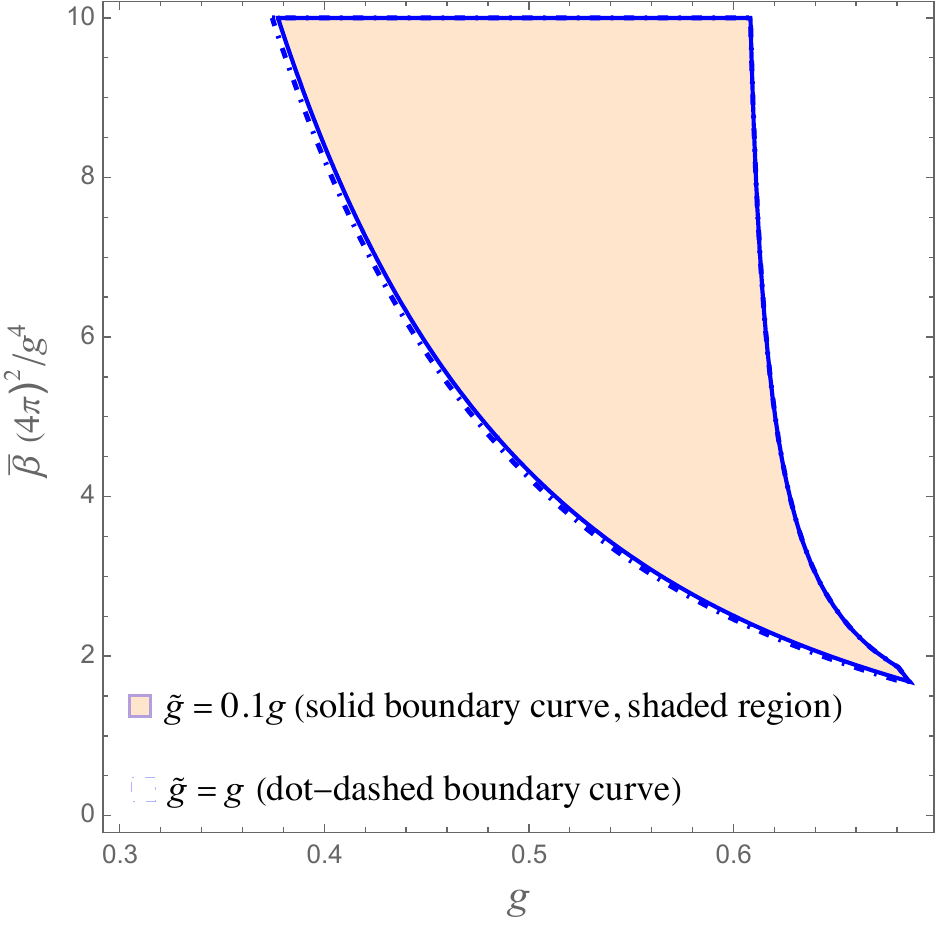} 
      \caption{\em Regions where $\Omega_{\rm GW}(f_{\rm peak})$ is above the sensitivities of LIGO-VIRGO O3 (left plot, where $\chi_0 = 2 \times 10^9$~GeV) and LISA (right plot, where $\chi_0 = 10^4$~GeV) computed at NLO in the supercool expansion. In both plots $g_*(T_r)=110$ and fast reheating is assumed. }\label{LIGO-LISA}
  \end{center}
\end{figure}

In Fig.~\ref{LIGO-LISA} we show the regions where $\Omega_{\rm GW}(f_{\rm peak})$ is above the sensitivities of LIGO-VIRGO O3 (left plot) and LISA (right plot) for two non-vanishing values of $\tilde g$. These plots should be compared respectively with the inset in Fig.~\ref{betaVT} and with the right plot in~\ref{LISA} (which have instead $\tilde g \approx 0$). Note that a value of $\tilde g$ much smaller than $g$ is possible: it can happen in models where there are several particles with comparable values of their couplings to $\chi$.
In Fig.~\ref{LIGO-LISA} we considered only values of $g$ and $\bar\beta$ such that $\epsilon$ (computed at LO) is below one.

One can then go ahead with this expansion. The next step, the next-to-next-to-leading order (NNLO), would consists in including terms of order $\epsilon \log\epsilon$ and $\epsilon$. This would require including the $x^2\log x$ and $x^2$ terms in the small-$x$ expansion of $J_B(x)$ and $J_F(x)$, Eqs.~(\ref{JBdef})-(\ref{JFdef}), and considering the field-dependent logarithm in~(\ref{logSplit}). Consequently also the $X\log X$ and $X$ terms  in the left-hand side of Eq.~(\ref{TnEqNLO0})
for the nucleation temperature $T_n$ (which needs to be corrected at NNLO) should be considered. In doing these improvements one would need to include extra parameters related to the fourth powers and the logarithms of the couplings of all fields to $\chi$. Because of the small numbers in front of $x^2\log x$ terms in Eqs.~(\ref{JBdef})-(\ref{JFdef}) 
and the fact that the terms $\log(\chi_b/T)-1/4$ in~(\ref{logSplit}) are at most of order $\log X/X< \epsilon \log\epsilon$ (relatively to the LO)  
one expects that their inclusion would generically lead to tiny corrections as long as $\epsilon$ is small.
A complete analysis of the NNLO and higher orders goes beyond this work and is left for future activities. 

 \section{Summary and conclusions}\label{Conclusions}
 
 Let us start this final section with a summary of the main original results of this paper.
 
\begin{itemize}
\item To begin with, in Sec.~\ref{theoretical framework} it has been shown  that in an arbitrary RSB scenario (where symmetries are broken and masses are generated mostly by perturbative radiative effects) the one-loop effective potential, which has been computed explicitly  and includes both quantum and thermal contributions, is always real as long as the classical potential is bounded from below. This an important starting point to  establish the validity of the loop expansion.
\item The PT of any RSB model is of first order when  analyzed with perturbative methods as shown in Sec.~\ref{1stPT}. However, in Sec.~\ref{supercool} it has then been proved that, for our purposes, one can trade the hypothesis of perturbativity with that of supercooling (together with, of course, that of small-enough couplings) to obtain that the symmetric and asymmetric configurations, $\chi=0$ and $\chi=\chi_0$, are always separated by a barrier for the relevant temperatures, $T\ll \chi_0$.
\item Sec.~\ref{supercool} contains other very important results. There it has been shown that if supercooling is large enough, namely if $\epsilon$ defined in~(\ref{CondConv}) is small, a model-independent description of the first-order phase transition is possible in terms of the following key parameters.
\begin{itemize}
\item $\chi_0$: the symmetry breaking scale 
\item $\bar\beta$: the beta function of the quartic coupling $\lambda_\chi$ of the flat-direction  field $\chi$, evaluated at the scale where $\lambda_\chi=0$. 
\item $g$: a sort of collective coupling of $\chi$ to all fields of the theory, which is precisely defined in Eq.~(\ref{M2g2def}). It is basically the square root of the sum of the squares of the couplings of $\chi$ to all fields.
\end{itemize}  
These parameters can be computed and the smallness  of $\epsilon$ can be checked once the model is specified.
As discussed in Sec.~\ref{supercool}, when the theory is perturbative and, thus, approximately scale invariant, a large amount of cooling is always present~\cite{Witten:1980ez}, and consequently the PT is strongly of first order, but here it has been found that, for the model-independent approach  to work, having a small $\epsilon$ is important\footnote{A non-perturbative way to realize a quasi-scale invariant scenario for PTs, on the other hand, is to use the AdS/CFT correspondence (see e.g.~\cite{Creminelli:2001th,Randall:2006py,Nardini:2007me, Konstandin:2011dr}). Also in these models supercooling  occurs  and the phase transition lasts a long time. The advantage of the perturbative scenario discussed here is the possibility of a model-independent description (at least for a large-enough supercooling), which has not (yet) been established at the non-perturbative level.}. 
\\   
 In the same section the validity of the one-loop approximation and the derivative expansion has been shown to follow from the hypothesis of supercooling, together with, of course, the requirement of small-enough couplings. This provides an extra motivation for considering the RSB scenario. In general, the gravitational contribution to the tunneling process turns out to be negligible for $\chi_0\ll M_P$. Moreover, we have shown that the tunneling process is always dominated by the time-independent bounce as long as $\epsilon$ is small.
 \item The description of the PT in terms of only $\chi_0$, $\bar\beta$ and $g$  is a leading order (LO) approximation in a small $\epsilon$ (supercool) expansion. The smaller $\epsilon$ the more accurate this approximation is; in general the errors associated with the LO approximation are of relative order not larger than $\sqrt{\epsilon}$ (see Sec.~\ref{impro}), but can even be smaller in specific setups. A next-to-leading (NLO) approximation can be obtained by including one more parameter, $\tilde g$, precisely defined in~(\ref{gtdef}). This parameter is basically the cube root of the sum of the cubes of the couplings of $\chi$ to all bosonic fields, so $\tilde g\leq g$. Just like $\chi_0$, $\bar\beta$ and $g$, the extra parameter $\tilde g$  can be computed once the model is specified.
  The NLO approximation should describe any RSB model modulo terms of relative order not larger than $\epsilon \log \epsilon$. One can improve even further the precision by going to the next-to-next-leading order (NNLO), by including additional parameters, as it is natural to expect. These results regarding the accuracy of the approximations and the supercool expansion are  provided in Sec.~\ref{impro}. In that section we have also computed the bounce action, a key ingredient to study the PT, as well as the nucleation temperature $T_n$ and the inverse duration of the PT, $\beta$, at NLO. 
 \item In Secs.~\ref{Gravitational Waves} and~\ref{impro} the corresponding GW spectrum is studied and compared with the experimental results and the expected sensitivities of current and future GW detectors, to find regions of the parameter space that are either already ruled out or can lead to a future detection. If reheating is fast, the spectrum  can be described at LO by $\chi_0$, $\bar\beta$ and $g$ only (with a weak dependence on the effective number of relativistic species $g_*$); in going to NLO one adds the extra parameter $\tilde g$. If reheating is not fast one should also include the reheating temperature and the corresponding Hubble rate among the independent parameters, but for a fast reheating these are dependent on the previously-mentioned parameters.  Interestingly, a large supercooling ($\epsilon$ small), which allows us to obtain a model-independent description, also generically increases the duration of the PT and makes it more likely for these GW signals to be observed in the future.
\item It is also important to mention that the LO and NLO approximations have been carried out analytically in terms of the key parameters, $\chi_0$, $\bar\beta$ and $g$, as well as $\tilde g$ for the NLO approximation.  This  allows  to easily compute the key quantities of PTs and GW spectra in a supercool expansion once
the model is specified.
\end{itemize}
To conclude, in this work we constructed a model-independent approach to RSB and the consequent PTs and GW spectra when enough supercooling occurs and we established the accuracy of such approach working at LO and NLO in the supercool expansion. The analytical formul\ae~produced in the paper are ready to be used in specific models and setups to avoid repeating the study of the PTs and GW spectra. 
  The detailed analysis at higher orders, starting from the NNLO, is another possible outlook for future work.

 \vspace{0.2cm}

 \subsubsection*{Acknowledgments}
I thank Giuliano Panico and Michele Redi for useful correspondence. This work has been partially supported by the grant DyConn from the University of Rome Tor Vergata.

 \vspace{1cm}
\footnotesize
\begin{multicols}{2}

\end{multicols}

\end{document}